\documentstyle[12pt,epsf,amstex]{article}
\def\fun#1#2{\lower3.6pt\vbox{\baselineskip0pt\lineskip.9pt
\ialign{$\mathsurround=0pt#1\hfil##\hfil$\crcr#2\crcr\sim\crcr}}}

\newcommand{\be}{\begin{eqnarray}}
\newcommand{\ee}{\end{eqnarray}}
\newcommand{\bd}{\begin{displaymath}}
\newcommand{\ed}{\end{displaymath}}
\newcommand{\ba}{\begin{array}}
\newcommand{\ea}{\end{array}}
\newcommand{\bt}{\begin{tabular}}
\newcommand{\et}{\end{tabular}}

\renewcommand{\theequation}{\thesection.\arabic{equation}}

\input epsf
\newcommand{\grpicture}[1]
{
    \begin{center}
        \epsfxsize=200pt
        \epsfysize=0pt
        \vspace{-5mm}
        \parbox{\epsfxsize}{\epsffile{#1.eps}}
        \vspace{5mm}
    \end{center}
}
\fontsize{12}{25}
\begin{document}

\vspace{.5cm}

\begin{center}
{\large
 Vacuum structure in supersymmetric Yang--Mills theories with any gauge group.}

\vspace{.5cm}

 {\bf V. G. Kac} \\
{\it IH\'ES, Le Bois--Marie, 35 route de Chartres, F-91440,
Bures-sur-Yvette, France \\
and \\
Departement of Mathematics, M.I.T., Cambridge, Massachusetts 02139, USA }\\ 
\ \ \\
and  {\bf A.V. Smilga}\\
{\it Universit\'e de Nantes, 2, Rue de la Houssini\`ere, BP 92208, F-44322,\\
Nantes CEDEX 3, France \\
and\\
ITEP, B. Cheremushkinskaya 25, Moscow 117218, Russia}

\end{center}

\vspace{.5cm}

\begin{abstract}
We consider the pure 
supersymmetric Yang--Mills theories placed on a small 3-dimensional spatial 
torus with higher orthogonal  and exceptional 
 gauge groups. The problem of constructing the quantum vacuum states is reduced
to a pure mathematical problem of classifying the flat connections on $T^3$.
The latter problem is equivalent to the problem of classification of commuting
triples of elements in a connected simply connected compact Lie group
which is solved in this paper. In particular,
we show that for higher orthogonal $SO(N), N \geq 7$, and for 
all exceptional groups
the moduli space of flat connections involves several distinct connected
components. The total number of vacuum states is given in all cases by the dual
Coxeter number of the group which agrees with the result obtained earlier
 with the instanton technique.
\end{abstract}

\section{Introduction}

It was shown by Witten long time ago \cite{IWit} that, in a pure $N=1$ 
supersymmetric gauge theory with any simple gauge group, the supersymmetry 
is not broken spontaneously. Placing the theory in a finite spatial box, 
the number of supersymmetric vacuum states [the Witten index ${\rm Tr} (-1)
^F$ ] was calculated to be  ${\rm Tr} (-1)^F = r + 1$ where $r$ is the rank 
of the gauge group. This result conforms with other estimates for 
 ${\rm Tr} (-1)^F$ for unitary and symplectic groups. For higher 
orthogonal and exceptional groups, it {\it disagrees}, 
however, with the general result 
\footnote{which follows e.g. from the counting of gluino zero modes on the 
instanton background and also from the analysis of weakly coupled theories
with additional matter supermultiplets \cite{IWit,IDyn}.}
  \be
 \label{IW}
 {\rm Tr} (-1)^F \ =\ h^\lor\ ,
 \ee
where $h^\lor$ is the dual Coxeter number of the group (see e.g. \cite{K}, 
Chapt. 
6; it coincides with the Casimir $T^aT^a$ in the adjoint representation when
a proper normalization is chosen.).
 For $SO(N \geq 7)$, $h^\lor = N-2 
 >  r+1$.
Also for exceptional groups $G_2, F_4, E_{6,7,8}$, the index (\ref{IW}) is 
larger  than Witten's original estimate (see Table 1).

\begin{table}

\begin{tabular}{||l|c|c|c|c|c||} \hline
group $G$& $G_2$ & $F_4$ & $E_6$ & $E_7$ & $E_8$ \\ \hline
r + 1    & 3 & 5 & 7 & 8 & 9 \\ \hline
 $h^\lor$ & 4&  9 & 12 & 18 & 30 \\ \hline
mismatch & 1& 4 & 5 & 10 & 21 \\ \hline
\end{tabular}

\caption{ Vacuum counting for  exceptional groups.}

\end{table}

This paradox persisting for more than 15 years has been recently resolved by 
Witten himself in the case of orthogonal groups \cite{Witnew}. 
He has found a flaw in his original arguments and 
shown that, for $SO(N \geq 7)$, vacuum moduli space is richer than it was 
thought before so that the {\it total} number  of quantum vacua is $N-2$ in 
accordance with the result (\ref{IW}).
In recent paper \cite{KRS}, this result was confirmed, and also the theory 
with the $G_2$ gauge group was analyzed where an extra vacuum state has been 
found. The aim of our paper is to develop a general method to construct the
 moduli 
space of flat connections for any gauge group including higher exceptional 
groups  $F_4, E_{6,7,8}$. As was anticipated in \cite{KRS}, the moduli space
for these groups involves not one and not two, but many distinct components 
(up to 12 components for $E_8$). The total number of quantum vacuum states 
coincides with $h^\lor$ in all cases.

Let us first recall briefly Witten's original reasoning. 
 \begin{itemize}
\item Put our theory on the spatial 3D torus and impose {\it periodic} 
boundary conditions on the gauge fields \footnote{For unitary groups, 
one can perform the counting also with 't Hooft twisted boundary conditions, 
but for the orthogonal and exceptional groups where the mismatch in the 
Witten index calculations was observed this method does not work.}. Choose 
the gauge $A_0^a = 0$. A classical vacuum is defined as a gauge field 
configuration $A_i^a(x,y,z)$ with zero field strength (a {\it flat 
connection} in mathematical language).
 \item For any flat periodic connection, we can pick out a particular point 
in our torus $(0,0,0) \equiv (L,0,0) \equiv \ldots$ and define a set of 
holonomies (Wilson loops along non--trivial cycles of the torus)
 \be
 \label{hol}
\Omega_1 \ &=&\ P \exp \left\{ i \int_0^L A_1(x,0,0)  dx \right\}\ ,
\nonumber \\ 
\Omega_2 \ &=&\ P \exp \left\{ i \int_0^L A_2(0,y,0)  dy \right\}\ ,
\\ 
\Omega_3 \ &=&\ P \exp \left\{ i \int_0^L A_3(0,0,z)  dz \right\}
\nonumber 
  \ee
( $A_i = A_i^a T^a$ where $T^a$ are the group generators in a given 
representation).
${\rm Tr} \{\Omega_i\}$ are invariant under periodic gauge transformations. 
 \item A necessary condition for the periodic connection to be flat is that 
all the 
holonomies (\ref{hol}) commute $[\Omega_i, \Omega_j] = 0$. For a 
{\it simply connected} group with $\pi_1(G) = 1$, it is also
a sufficient condition.
  \item A sufficient condition for the group matrices to commute is that
their {\it logarithms} belong to a Cartan subalgebra of the corresponding 
Lie algebra. For unitary and simplectic groups, this happens to be also a 
necessary 
condition. In other words, any set of commuting group matrices $\Omega_i$ 
with $[\Omega_i, \Omega_j] = 0$ can be presented in the form
 \be
\label{Cart}
\Omega_i \ =\ \exp\{iC_i\} ,\ \ \ \ [C_i, C_j] = 0 \ .
 \ee
A flat connection with the holonomies $\Omega_i$ is then just $A_i = C_i/L$.
The moduli space of all such connections presents ( up to factorization
over the discrete Weyl group) the product $T_C \times T_C
\times T_C$ where $T_C$ is the Cartan torus whose dimension coincides with 
the rank $r$ of the group.

\item  The Witten's original assumption which 
came out not to be true is that this is also the case for all other groups.
Assuming this, Witten constructed an effective Born--Oppenheimer 
hamiltonian for the slow variables $A_i^a$ where the index $a$ runs only over 
the Cartan subalgebra. It involves $3r$ bosonic degrees 
of freedom  and their fermionic counterparts $\lambda_\alpha^a$ ($\alpha = 1,2$
so that the gluino field $\lambda_\alpha^a$ is the Weyl spinor). 
Imposing further the condition of the Weyl symmetry (a remnant of the 
original gauge symmetry) for the eigenstates of 
this hamiltonian, one finds $r+1$ supersymmetric quantum vacuum states:
\footnote{See Appendix B  for a rigorous proof.}
\be
\label{vacT}
\Psi(A_i^a, \lambda_\alpha^a) \ \sim\  1, \epsilon^{\alpha\beta} \gamma^{ab}
\lambda_\alpha^a \lambda_\beta^b, \ldots ,  \left( \epsilon^{\alpha\beta} 
\gamma^{ab} \lambda_\alpha^a \lambda_\beta^b\right)^r \ ,
\ee
where $\gamma^{ab} = \delta^{ab}$ if an orthonormal basis in the Cartan 
subalgebra is chosen.
 \end{itemize}
However, for the groups $ Spin(N)$
\footnote{$ Spin (N)$ is the simply connected universal covering of $SO(N)$. It is
represented by the matrices $\exp\{\omega_{\mu\nu} \gamma_\mu \gamma_\nu/2 \}$
where $\gamma_\mu$ are the gamma--matrices of the corresponding dimension 
forming the Clifford algebra $\gamma_\mu \gamma_\nu + \gamma_\nu \gamma_\mu = 
2\delta_{\mu\nu}$, $\mu = 1, \ldots, N$. We have already mentioned and will see
later that the condition $\pi_1(G) = 0$ is very important in the whole 
analysis. It what follows, we will discuss mostly the groups  $Spin (N)$ 
rather than their orthogonal counterparts.  }
and for all  exceptional groups, there are some  triples of the group
elements $(\Omega_1, \Omega_2, \Omega_3)$ which commute with each other but 
which cannot be conjugated (gauge--transformed) to the maximal Cartan torus.
\footnote{This fact has actually been noticed by topologists long time ago. 
See e.g. Refs.
\cite{serr,SS} where some examples of such non--trivial triples were 
constructed.}

 In the case of  {\it Spin}(7), there is a unique 
[up to conjugation $\Omega_i \to g^{-1} \Omega_i g,\ \ g \in Spin(7)$ ]
 non--trivial
triple. It can be chosen in the form \cite{SS,Witnew}
\be
\label{Spin7}
 \Omega_1 \ =\ \exp\left\{ \frac \pi 2 (\gamma_1 \gamma_2 + \gamma_3 \gamma_4)
\right\} = \gamma_1\gamma_2\gamma_3\gamma_4 \ , \nonumber \\ \ 
 \Omega_2 \ =\ \exp\left\{ \frac \pi 2 (\gamma_1 \gamma_2 + \gamma_5 \gamma_6)
\right\} = \gamma_1\gamma_2\gamma_5\gamma_6 \ , \nonumber \\\ 
 \Omega_3 \ =\ \exp\left\{ \frac \pi 2 (\gamma_1 \gamma_3 + \gamma_5 \gamma_7)
\right\} = \gamma_1\gamma_3\gamma_5\gamma_7  \ .
  \ee
Obviously, $[\Omega_i, \Omega_j] = 0$.
On the other hand, as we will shortly be convinced, the triple (\ref{Spin7}) 
cannot be conjugated to the maximal torus. This isolated triple corresponds to
an isolated flat connection on $T^3$ which brings about the extra quantum 
vacuum state. Thereby,
\be
 {\rm Tr} (-1)^F \ =\ r_{Spin(7)} + 1 + 1 = 5 = h^\lor_{Spin(7)} \ . 
 \ee
 For {\it Spin}(8) and $G_2$, there is also only one extra isolated triple and
one extra vacuum state. For {\it Spin}($N \geq 9$), there is a family of such 
triples with each element lying on the ''small''  torus which can be
interpreted as the Cartan torus of the {\it centralizer} 
\footnote{The centralizer of the element $g$ of the group $G$ is defined as a 
subgroup  of $G$ involving all elements $h$ which commute with $g$. The 
centralizer of a triple is the subgroup of $G$ commuting with each element of 
the triple.} of the triple (\ref{Spin7}). The continuous part of
the latter   is  $ Spin(N-7)$.
 This brings about $r_{{ Spin}(N-7)} + 1$ extra vacuum states. The total 
counting is
\footnote{
 In the case $N=9$, there is a subtlety : the Witten's index associated with
the abelian $Spin(2) = U(1)$ gauge group turns out to be zero, not 
$r_{U(1)} + 1 = 2$.
 Still, the
counting (\ref{count}) is correct due to the presence of certain discrete 
factors in the centralizer \cite{Witnew,KRS} 
which will be discussed in more details
in Appendix C. Moreover, we will show 
in Appendix A that, for  odd $N$, one can choose 
the triple in such a way that the continuous part of its centralizer would be 
$Spin(N - 6)$ rather 
than $ Spin(N-7)$. For $N=9$ that gives $r_{Spin(3)} + 1 = 2$ extra vacuum
states. Also  for
$N = 11,13,\ldots$, the  counting (\ref{count}) holds due to the fact 
that the ranks of $Spin(N - 6)$ and $Spin(N - 7)$ coincide for odd $N$.}
  \be
\label{count}
 {\rm Tr} (-1)^F \ =\ r_{{ Spin}(N)} + 1 + r_{{Spin}(N-7)} + 1
 = N-2 = h^\lor_{Spin(N)} \ . 
\ee
Before proceeding further, let us make a simple remark. For a group which is
not simply connected, there are always non--trivial commuting {\it pairs} of the
elements which cannot be conjugated on the maximal torus. Consider for example
$SO(3)$. The elements diag(-1, -1, 1) and diag(-1, 1, -1) obviously commute, 
but their logarithms proportional to the generators of the correponding
rotations $\sim T_3$ and $\sim T_2$ do not. In the covering $SU(2)$ group, the
corresponding elements $i\sigma_3$ and $i\sigma_2$ {\it anticommute}. Such a 
pair of the elements of $SU(2)$ is usually called the Heisenberg pair. 
Generally, we will call the Heisenberg pair a pair of the elements 
$\tilde b,\tilde c$ which satisfy the relations
 \be
\label{Heis}
\tilde b \tilde c \ =\ z \tilde c \tilde b
 \ee
where $z$ is the element of the center of the group.
An example of the Heisenberg pair for $SU(3)$ with $z = e^{2 \pi i/3} I$ is
\be
\label{HeisSU3}
\tilde b \ =\ \left( \begin{array}{ccc}
1& 0&0 \\
0 & e^{2 \pi i/3}&0 \\
0&0& e^{-2 \pi i/3}
\end{array} \right), \ \ 
\tilde c \ = \ \left( \begin{array}{ccc}
0& 0&1 \\
1 & 0&0 \\
0&1& 0
\end{array} \right) \ .
\ee
A permutation of $\tilde b$ and $\tilde c$ gives the pair with 
$z = e^{-2 \pi i/3} I$.
As we will prove later, all Heisenberg pairs in $SU(3)$ are equivalent  by 
conjugation to (\ref{HeisSU3}) or its permutation.
 Heisenberg pairs will play the central role in the whole analysis.

For a simply connected group, a pair of commuting elements $g,h$ can always be 
conjugated to the maximal torus. That follows from the fact that {\it one} 
element can always be put on the torus and from the so called Bott's theorem:
the centralizer $G_g$ 
 of any element $g$ in a connected simply connected compact Lie 
group $G$ is always connected. We first put
$g$ on the torus of $G$ and then conjugate $h$ to the torus of the centralizer
which coincides with the large torus (obviously, once $g$
is put on the torus, any element of the torus
belongs to the centralizer).

Consider an arbitrary simply connected group $G$. For a commuting triple of 
elements $g_{1,2,3}$, the elements $g_2, g_3$ belong to the centralizer 
$G_{g_1}$
of $g_1$. Now if $G_{g_1}$ presents a product of a  simply connected group and
some number of $U(1)$ factors as is usually the case 
(e.g. the centralizer of an element of $SU(3)$ can be $SU(3)$, $SU(2) \times
U(1)$ or $[U(1)]^2$ ), one can put $g_2$ and $g_3$ 
on the
torus of $G_{g_1}$ and hence the whole triple on the torus of $G$. But if 
$G_{g_1}$ involves a factor whose fundamental group is finite, one 
cannot do it in a general case.

One can prove (we will do it later) that the centralizer  of any element  of a 
unitary or a simplectic group is always simply connected up to some number
of  $U(1)$ factors and hence these 
groups  do not admit non--trivial triples. A central statement is that it is 
not so for higher orthogonal and exceptional groups. And this is the origin of 
non--trivial triples.

Look again at the triple (\ref{Spin7}). The element $\Omega_1$ commutes with
6 generators $i \gamma_1 \gamma_2/2\ , \ldots $ of $Spin(4)$ acting on the 
subspace
(1234) and with 3 generators of $Spin(3)$ acting on the subspace (567). Hence,
the centralizer of $\Omega_1$ seems to be $Spin(4) \times Spin(3)$. This is 
not, however, quite so: the whole product  $Spin(4) \times Spin(3)$ is not a
subgroup of $Spin(7)$. Indeed, both $Spin(4)$ and $Spin(3)$ have non--trivial
center. The center of $Spin(3) \equiv SU(2)$ is ${\bf Z}_2$, a non--trivial center
element corresponding to rotation by $2\pi$ in, say, (56) - plane.
 The center of
$Spin(4) \equiv SU(2) \times SU(2)$ is ${\bf Z}_2 \times {\bf Z}_2$. It has a ''diagonal``
element  presenting the product of non--trivial center elements in each $SU(2)$
factor: 
  \be
 z \ =\ \exp \left\{ \frac \pi 2 (\gamma_1 \gamma_2 - \gamma_3 
\gamma_4 ) \right\} 
\exp \left\{ \frac \pi 2 (\gamma_1 \gamma_2 + \gamma_3 
\gamma_4 ) \right\} \ =\ \exp \left\{  \pi  \gamma_1 \gamma_2  \right\} \ .
 \ee
Note now that $\exp \left\{  \pi  \gamma_1 \gamma_2  \right\} =
\exp \left\{  \pi  \gamma_5 \gamma_6  \right\}\ = -1$ in the full
 $Spin(7)$ group. Hence the true centralizer of $\Omega_1$ is 
 $[Spin(4) \times Spin(3)]/{\bf Z}_2 \equiv [SU(2)]^3/{\bf Z}_2$ where the factorization 
is
done over the common center of  $Spin(3)$ and
  of each $SU(2)$ factor in $Spin(4)$. The matrices $\Omega_2$ and
$\Omega_3$ correspond to the elements $(i\sigma_3 , i\sigma_3 ,
i\sigma_3 )$ and $(i\sigma_2 , i\sigma_2 ,
i\sigma_2 )$ in $[SU(2)]^3$, i.e. they form a Heisenberg pair in each $SU(2)$
factor. They still commute in the centralizer and in $Spin(7)$ exactly because
the factorization over the common center should be done. The elements
$\Omega_2, \ \Omega_3$ cannot be conjugated to the maximal torus in the 
centralizer and hence the whole triple (\ref{Spin7}) cannot be conjugated to
the maximal torus in $Spin(7)$.

What is the underlying reason for the fact that non--trivial triples exist
in $Spin(7)$, in higher $Spin(N)$ and in exceptional groups, but do not exist
in unitary and simplectic groups ? To understand it, one has to consider the
{\it Dynkin diagram} for the simple roots of the corresponding algebras. 

The several following paragraphs are addressed for the reader who is not an
expert in the compact Lie group theory with a hope to give him a feeling what 
is going
on here. An expert can skip it and continue reading from the beginning of
 the next section.

\begin{figure}
\grpicture{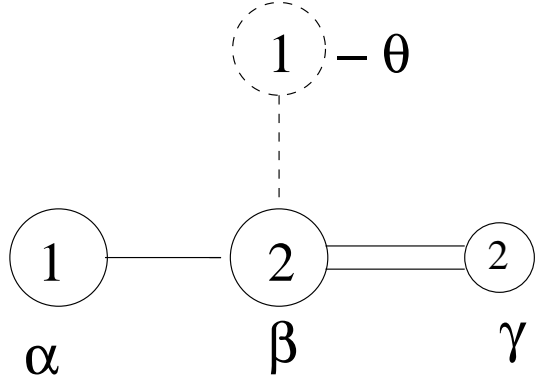}
\caption{Dynkin diagram for $Spin(7)$ with its Dynkin labels.}
\label{Dyn7}
\end{figure}

The Dynkin diagram for $Spin(7)$ is depicted in Fig. \ref{Dyn7} . 
$\alpha$, $\beta$,
and $\gamma$ are the simple roots. 
The term ''root'' means that the corresponding generators 
$T_{\alpha,\beta,\gamma}$ are eigenvectors of the (complex) Lie algebra of
$Spin(7)$  with respect to its Cartan subalgebra. In other words, they satisfy
the relations
 \be
\label{rootdef}
[e_i, T_\alpha]& =& \alpha_i T_\alpha, \  \ 
[e_i, T_\beta] = \beta_i T_\beta, \  \ 
[e_i, T_\gamma] = \gamma_i T_\gamma \ ,
 \ee
where $e_i$ is the normalized orthogonal basis in the Cartan subalgebra. A 
convenient choice is $e_1 = T_{12}, \
e_2 = T_{34},\ e_3 = T_{56}$ where $T_{ij} = i\gamma_i \gamma_j/2$ are the 
generators of $Spin(7)$. 
 In that case
  \be
 \label{genroot}
T_\alpha \ &=& \ \frac 12  [T_{13} + T_{24} - i(T_{23} + T_{41})] \ , \ 
T_\beta \ = \  \frac 12 [T_{35} + T_{46} - i(T_{45} + T_{63})] \nonumber \\
T_\gamma \ &=& \ T_{57}  - iT_{67} \ ,
 \ee
[$T_{\alpha, \beta, \gamma}$ multiplied by any factor would also satisfy the
commutation relations (\ref{rootdef}), but the normalization chosen in
Eq.(\ref{genroot}) is the most convenient one].
The root vectors are
$\alpha_i = (1,-1,0),\ \beta_i = (0,1,-1)$ and $\gamma_i = (0,0,1)$.
One can define the scalar product $\langle .,.\rangle$ in a natural way. 
The squares of the 
lengths of the roots are equal to 2 for 
$\alpha, \beta$ and to 1 for $\gamma$. Thereby, $\gamma$ is a ''short root'' 
and it is denoted with the small circle. A single line between the roots 
$\alpha, \beta$ means that $\langle\alpha, \beta\rangle  = -1$ so that the 
angle between
these roots $\theta_{\alpha, \beta} = \arccos \left[\langle\alpha, 
\beta\rangle 
/\sqrt{\langle\alpha,
 \alpha\rangle \langle\beta, \beta\rangle }\right] = 120^o $. The double line
 between $\beta$ and 
$\gamma$ means that $\theta_{\gamma, \beta} = 135^o$. The absence of the line
between $\alpha$ and $\gamma$ means that these roots are orthogonal and 
the corresponding generators in the Lie algebra commute. Any other positive 
\footnote{ For any such root $\kappa$, there is also a negative root
$ - \kappa$. The commutator $[T_\kappa, T_{-\kappa}]$ lies in the
Cartan subalgebra. In the following, we will need also the notion of {\it 
coroot} which is defined as the element $\kappa^\lor$ of the Cartan 
subalgebra proportional
to [with the normalization choice as in Eq.(\ref{genroot}), just equal to] 
$[T_\kappa, T_{-\kappa}]$ such that  $[\kappa^\lor, T_{\pm \kappa}] =
\pm 2 T_{\pm \kappa}$. In other words, the generators 
$\kappa^\lor, T_{\pm \kappa}$ form the $A_1$ [or 
$su(2)$] subalgebra in the same way as the matrices $\sigma_3, \sigma_\pm$ do.
For any coroot $\kappa^\lor$, the property
$\exp\{2\pi i \kappa^\lor\} = 1$ holds.
The ratio of the lengths of some coroots $\kappa^\lor,\ \rho^\lor$ is always
inverse compared to the ratio of the lengths of the corresponding roots
 $\kappa,\ \rho$ (so that a short root corresponds to a long coroot and the 
other way round). 
 For the $Spin(7)$ group, the coroots
corresponding to the simple roots $(\alpha\beta\gamma)$
are $\alpha^\lor = T_{12} - T_{34},\  \beta^\lor = T_{34} - T_{56},\ 
\gamma^\lor = 2T_{56}$.}
root of the algebra can be presented as a linear combination
of the simple roots
$a_\alpha \alpha + a_\beta \beta + a_\gamma \gamma$ with positive 
integer $a_{\alpha,
\beta, \gamma}$. The sum $a_\alpha + a_\beta + a_\gamma$ is called the weight
of the root. The root with the highest weight $\theta = \alpha + 2\beta + 
2\gamma$
[the corresponding coroot $\theta^\lor$ is  $T_{12} + T_{34}$] plays a special
 role. It (or 
rather the corresponding
negative root $-\theta$ ) is denoted by the dashed circle in Fig. \ref{Dyn7}. 
The {\it 
Dynkin labels} $a_i = (1,2,2)$ seen on the diagram are just the coefficients of
the expansion of the highest root via the simple roots. It is convenient to
ascribe the label 1 for the negative highest root. We see that 
$-\theta$ has  length 2, is orthogonal to $\alpha$ and $\gamma$, and form the
angle $120^o$ with the root $\beta$.

We will prove later that whenever the algebra involve a simple root with the
property $a^\lor_i = a_i \langle\alpha_i, \alpha_i\rangle /2 \ > 1$, an element
 of the group exist
whose centralizer, with its 
$U(1)$ factors being stripped off,
(that is called the {\it semi-simple part} of the centralizer) 
 is not simply connected. And if not, then not. That is why 
such {\it exceptional} elements are absent for unitary and simplectic groups 
where $a^\lor_i = 1$ for all simple roots.

For $Spin(7)$, $a^\lor_\alpha = a^\lor_\gamma = 1$ and $a^\lor_\beta = 2$. 
Thereby,  the root $\beta$ provides us with the element whose centralizer 
is simply connected (no $U(1)$ factors here and the centralizer coincides with
its semi--simple part).  The explicit form of this element is
  \be
\label{gom}
\sigma \ = \ \exp\{\pi i \omega_\beta\} \ ,
 \ee
where $\omega_\beta$ is the {\it fundamental coweight} --- the element of the 
Cartan algebra satisfying the property $[\omega_\beta, T_\alpha]  =
[\omega_\beta, T_\gamma]  = 0,\ \ [\omega_\beta, T_\beta] 
  = T_\beta$. Indeed, in our case,
$\omega_\beta = \theta^\lor$ 
and $\sigma$ coincides with $\Omega_1$ in Eq. (\ref{Spin7}). 
The centralizer of $\sigma$ has been constructed explicitly above, but it can
 also
be calculated on the basis of the Dynkin diagram. Obviously, $\sigma$ 
commutes with
the generators corresponding to the simple roots $\alpha,\ \gamma$. But it 
also commutes with the generator $T_\theta$ corresponding to the highest
root $\theta$. That follows from the fact  that $[\omega_\beta, T_\theta] = 
 2T_\theta$. Thereby, the 
centralizer of $\sigma$
is formed by the commuting generators $T_\alpha, T_\gamma, T_\theta$ and is 
hence $[SU(2)]^3$.
The fact that this product has to be factorized over the common center ${\bf Z}_2$
is not seen right away  now, but we have seen it explicitly for $Spin(7)$ and
it will be proven  in a general manner in the next section.

\begin{table}
\begin{center}
        \epsfxsize=400pt
        \epsfysize=0pt
        \vspace{-5mm}
        \parbox{\epsfxsize}{\epsffile{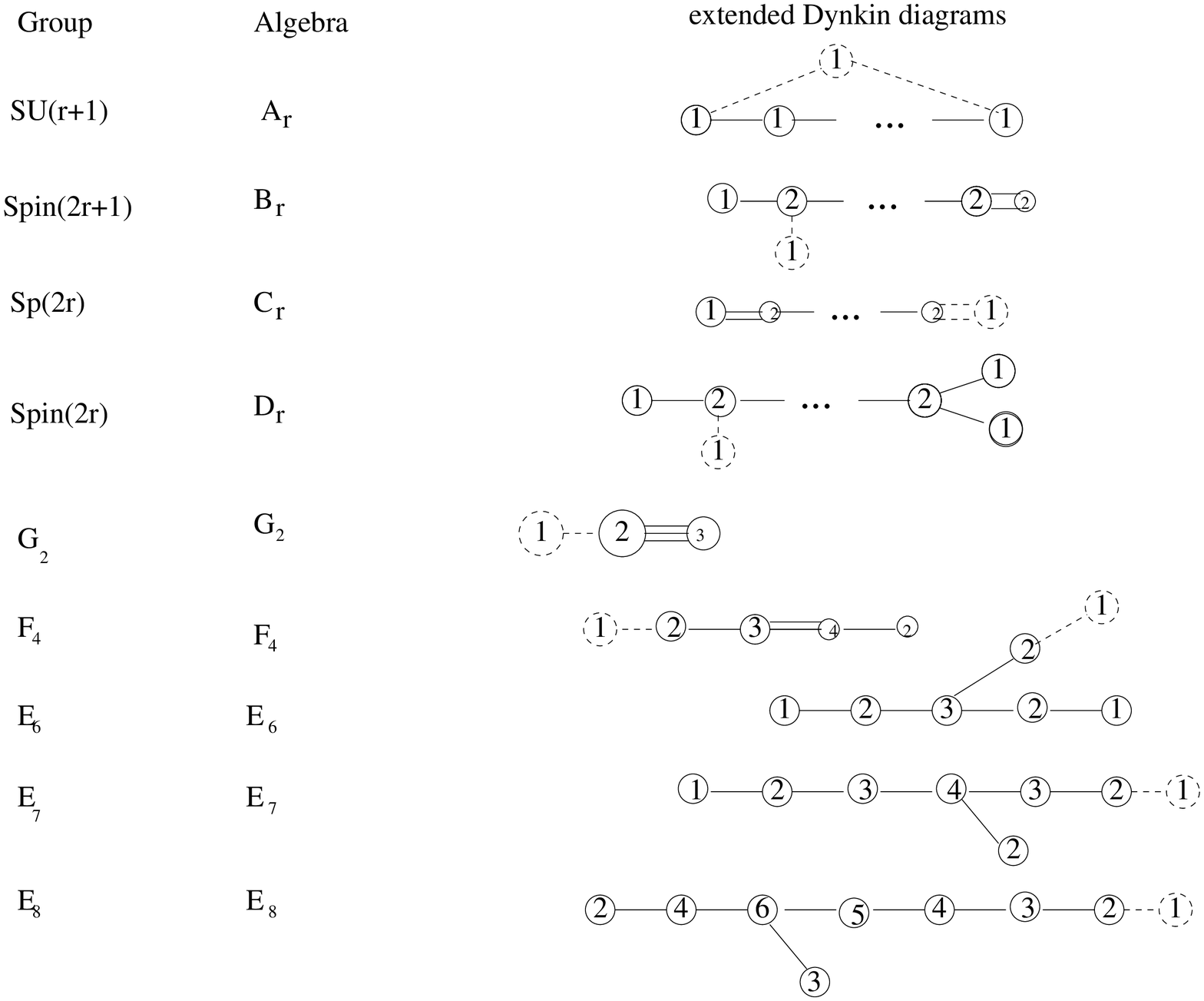}}
        \vspace{5mm}
    \end{center}
\caption{Dynkin diagrams and Dynkin labels for all compact Lie groups. Smaller
circles denote short simple roots.}
\label{Dynall}
\end{table}

Consider now  higher  groups $Spin(N \geq 9)$. As is seen from  
Table  \ref{Dynall}, the corresponding Dynkin diagrams involve several 
  nodes with   $a^\lor_i = 2$.
Each element $\sigma_j = \exp\{\pi i \omega_j\}$ has a 
 centralizer which is semi-simple and is not simply connected. 
Moreover, in this case we have a family of the elements
  \be
\label{famSpin} 
 \sigma_{\vec{s}} \ = \ 
\exp \left\{ 2\pi i \sum_j s_j \omega_j   \right\}
 \ee
where $s_i$ are some  real numbers with $2\sum_i s_i = 1$ and the sum 
runs over all the nodes with 
$a^\lor_i = 2$. If all $s_i$ are nonzero, the element (\ref{famSpin})
 commutes with
two (for odd $N$) and three (for even $N$) simple  roots with $a^\lor_i = 1$
  and with the highest root $\theta$. In this 
general case, the centralizer of the element (\ref{famSpin}) is $[SU(2)]^3/{\bf Z}_2
\times [U(1)]^{r-3}$
for  $N = 2r + 1$ and   $[SU(2)]^4/{\bf Z}_2 \times [U(1)]^{r-4}$ for  $N = 2r$. 
The  family (\ref{famSpin})
spans a $d$--dimensional torus in $Spin(N)$ ($d = r -3$ or $r - 4$
depending on whether $N$ is odd or even. It coincides with the Cartan
torus of a subgroup $Spin(N-7)$ (or $Spin(N - 6)$ for odd $N$) as is not 
difficult to see when substituting
the explicit expressions for the fundamental weights: 
$\omega_1 = i(\gamma_1 \gamma_2 +
\gamma_3 \gamma_4)/2, \ \omega_2 = i(\gamma_1 \gamma_2 +
\gamma_3 \gamma_4 + \gamma_5 \gamma_6)/2$, etc.

We will see later that one can restrict all $s_i$ in Eq. (\ref{famSpin})
to be positive and the elements (\ref{famSpin}) involving some negative
$s_i$ can be reduced by conjugation to the same canonical form (\ref{famSpin})
with all $s_i$ positive. Thereby, the moduli space of exceptional elements
(the space of exceptional gauge orbits) presents the torus $T^d = [U(1)]^d$ 
factorized
over the action of a certain discrete group which plays the role of the Weyl
group for this small torus.
For each such element (including also the case when some of $s_i$ vanish), the
moduli space of nonequivalent Heisenberg pairs in the centralizer presents
a subset  of
$T^{d} \times T^{d}$ ( or rather its quotient by a finite group). Thereby, the
 moduli space of all nonequivalent
triples lies in  $T^{d} \times T^{d} \times T^{d}$ in agreement with the 
previous results.

As is seen from the Table \ref{Dynall}, the exceptional groups also 
involve a 
number of nodes
with $a^\lor_i > 1$. For example, $G_2$ involves only one such node with 
$a^\lor_i = 2$
which gives rise to only one non--trivial isolated triple. Other exceptional
groups involve several such nodes, and the moduli space is somewhat 
more complicated but, as we will see, it can be rather easily constructed just
by the  form of the corresponding Dynkin diagram.

The  three following sections present the hard core of the paper. In the next 
section,
after some preliminary remarks, we establish the families of the exceptional
elements whose centralizer is not simply connected. The structure of the 
Heisenberg pairs in these centralizers is the subject of Sect. 3. In Sect. 4 we
construct the moduli space of all commuting triples. In Sect. 5, we 
spell out 
the results  for higher exceptional groups in a more explicit form.
In Appendix A, we calculate the centralizers of the Heisenberg pairs and of
non--trivial commuting triples for all gauge groups. In Appendix B, we give
a rigourous proof that each distinct component in the moduli space of triples
associated with the torus $T^d \times T^d \times T^d $ gives exactly $d + 1$
vacuum states.  Appendix C is devoted to calculation of one particular object
--- the group $\Gamma_{m}$ which is defined as the intersection of the 
relevant 
Cartan sub--torus with its centralizer (this group affects the global structure
of the thriples' moduli space). We also take the opportunity there
 to write down some further explicit illustrative formulae.

\section{Conjugacy classes and centralizers in compact Lie groups.}
\setcounter{equation}0

Let $G$ be a compact Lie group and let $Lie\ G$ be its Lie
 algebra. Fix
a $G$--invariant symmetric positive definite bilinear form $\langle .,.
\rangle $ on
$i\ Lie\ G$. Choose a maximal torus $T$ in $G$ and let ${\mathfrak h} = \ i\ 
Lie\ T$.
The bilinear form $\langle .,.\rangle $ defines an isomorphism of real vector
 spaces
$\nu: {\mathfrak h} \to {\mathfrak h}^*$ given by $h \to \nu(h):\ 
\nu(h) (h') = \ \langle h, h'\rangle , \ h,h' \in {\mathfrak h}$,
and hence a bilinear form $\langle.,.\rangle $ on ${\mathfrak h}^*$. Given a 
non-zero vector
$\alpha \in {\mathfrak h}^*$, the vector ${  \alpha^\lor} = 2\nu^{-1}(\alpha)
 /\langle\alpha,
\alpha\rangle  \in {\mathfrak h}$ is independent of the choice of 
$\langle.,.\rangle $.

Let $\Delta \subset {\mathfrak h}^*$ be the set of (non-zero) roots of $Lie\
 G$ and let
$\Delta^\lor = \{\alpha^\lor |\alpha \in \Delta\} \subset {\mathfrak h}$ be 
the set
of coroots. Denote by $Q(G)$ and $Q^\lor(G)$ the {\bf Z}--span of the sets
$\Delta$ and $\Delta^\lor$ , respectively. These are called the root and coroot
lattices and actually depend only on $Lie\ G$.
 Introduce the coweight
lattice
 \be
\label{coweight}
P^\lor (G) \ =\ \{ h \in {\mathfrak h} | e^{2\pi i h} = 1\} \ .
\ee
This lattice contains the lattice $Q^\lor (G)$. It is well known that if $G$
is a connected compact Lie group, then its fundamental group is given by the 
following formula
\be
\label{pi1}
\pi_1(G) \ =\ P^\lor(G)/Q^\lor(G) \ .
 \ee

Suppose now that $G$ is a connected simply connected (almost) simple
compact Lie group of rank $r$. Choose a set of simple roots
$\Pi = \{\alpha_1, \ldots, \alpha_r\} \subset \Delta$, then
$\Pi^\lor = \{\alpha_1^\lor, \ldots, \alpha_r^\lor\} \subset \Delta^\lor$
is a set of simple coroots. Let
  \be
\alpha_0 = \ -\theta = -\sum_{j=1}^r a_j \alpha_j, \ \ \ \ 
\alpha_0^\lor = \ -\sum_{j=1}^r a_j^\lor \alpha_j^\lor 
 \ee
be the lowest root [i.e. $\alpha_0 - \alpha_j$
 are not roots  for all $j = 1,\ldots, r$] and its coroot.
 The numbers $a_j$ and $a_j^\lor$ are (well-known) positive
integers. 
Recall that $G$ of type $A$--$D$--$E$ is called {\it simply laced} (or
1--laced),  $G$ of type $B$--$C$--$F$ is called {\it 2--laced} and the group
$G_2$ is {\it 3--laced}.
If  $\alpha_j$ is a long
root, one has: $a_j^\lor = a_j$ (that happens for  all $j$ in
the $A-D-E$ cases). 
 If  $\alpha_j$ is a short
root, one has $a_j^\lor = a_j/l$ if $G$ is $l$--laced. 
Let $a_0^\lor = 
a_0 = 1$.
The set of vectors $\{\alpha_0, \alpha_1, \ldots, \alpha_r\}$ along
with the integers $a_j$ are depicted 
by the extended Dynkin diagrams $\hat D (G)$ which are listed in  
Table   \ref{Dynall}. The Dynkin diagram $D(G)$ of $G$ is obtained from
$\hat D(G)$ by removing $\alpha_0$--th node.

Define fundamental coweights $\omega_1, \ldots, \omega_r \ \in {\mathfrak h}$
by $\alpha_i(\omega_j) = \delta_{ij}, \ i,j = 1,\ldots, r$. Note that their
{\bf Z}--span is the coweight lattice of the adjoint group $Ad\ G$. 

{\bf Theorem 1}. 
Let $G$ be a connected simply connected (almost) simple
compact Lie group. For a set of $r +1$ real numbers $\vec{s} = 
(s_0, s_1, \ldots, s_r)$ such that
 \be
\label{conds}
s_j \geq 0,\ j = 0, \ldots, r;\ \ \ \sum_{j=0}^r a_j s_j = 1
 \ee
consider the element 
  \be
\label{excel}
\sigma_{\vec{s}} = \exp \left\{ 2\pi i  
\sum_{j=1}^r s_j \omega_j \right\} \ \in \ T\ .
 \ee

{\it (a)} Any element of ${ G}$ can be conjugated to a unique element
$\sigma_{\vec{s}}$.

{\it (b)} The centralizer of $\sigma_{\vec{s}}$ ( in  ${ G}$) is a 
connected 
compact Lie group which is a product of $[U(1)]^{n-1}$
where $n$ is the number of non-zero $s_i$, and a connected semi-simple group 
whose Dynkin diagram is obtained from the extended Dynkin diagram  
$\hat D({ G})$ by removing 
the nodes $i$ for which $s_i \neq 0$.
  
{\it c)} The fundamental group of the centralizer
of $\sigma_{\vec{s}}$ is isomorphic to a direct product
of ${\bf Z}^{n-1}$ and a cyclic group of order $a_{\vec{s}}$ where
  \be
 a_{\vec{s}} \ =\ \gcd \{ a_j^\lor | s_j \neq 0, j = 0,\ldots, r \} \ .
 \ee
In particular, the fundamental group of the semi-simple part of the 
 centralizer of $\sigma_{\vec{s}}$ is the cyclic group of order $a_{\vec{s}}$.

{\bf Corollary 1}.
The connected subgroup $H$ of $G$  corresponding to a 
subdiagram of $D(G)$ is simply connected.

{\bf Definitions}. If $g \in G$ is conjugate
to $\sigma_{\vec{s}}$, the real numbers $s_0, \ldots , s_r$ are called 
{\it coordinates} of $g$.  We will call the element $g$ 
  $m$--{\it exceptional}  whenever $m =  a_{\vec{s}} > 1$. The set of 
all possible coordinates  for the $m$--{\it exceptional} elements will be
called {\it fundamental alcove} of type $m$.
A positive integer $m$ is called $G$--{\it exceptional} if 
it divides one of the $a_j^\lor$ ($j = 0, \ldots, r$). 

A look at  Table  \ref{Dynall} gives the following possibilities for 
$G$- exceptional integers:
 \be
SU(N), \ Sp(2N)&:&\ \ \ 1 \nonumber \\ 
Spin(N), \ G_2 &:& \ \ \ 1,2 \nonumber \\   
F_4, \ E_6 &:& \ \ \ 1,2,3 \nonumber \\ 
E_7 &:& \ \ \ 1,2,3,4 \nonumber \\ 
E_8 &:& \ \ \ 1,2,3,4,5,6 
  \ee

{\bf Proof of Theorem 1}.
Statement {\it a)} is well known. In this form for finite order elements it can
be found e.g. in Ref.\cite{K}, Chapter 8.
The connectedness of the centralizer of any element of $G$ is usually 
attributed to Bott (see e.g. \cite{SS}, \S3). A proof of the second part of
{\it b)} can also be found in \cite{K}, Chapter 8.

Finally, the proof of {\it c)} is based on the formula (\ref{pi1}). If $\tilde G$ is
a subgroup of $G$ containing $T$, then $P^\lor(\tilde G) = P^\lor (G) =
Q^\lor(G)$ and
$Q^\lor(\tilde G) \subset Q^\lor (G)$, hence
  \be
 \pi_1(\tilde G) \ =\ Q^\lor(G)/Q^\lor(\tilde G)
 \ee
Note that by {\it b)}, $Q^\lor(G_{\sigma_{\vec{s}}})$ is spanned over ${\bf Z}$
by those $\alpha_j^\lor$ for which $s_j = 0$ ($j = 0,\ldots,r$).
The proof of {\it c)} now is completed by the following simple lemma.

{\bf Lemma 1}.
Let $Q = \oplus_{i=1}^r ${\bf Z}$ \alpha_i $ be a free abelian group of rank 
$r$ and let $\alpha_0 = \sum_{i=1}^r a_i \alpha_i$ be a non-zero element of 
$Q$. Let $I$ be a subset of $\{0,1,\ldots, r\}$ and let 
$Q_I = \oplus_{i \in I} {\bf Z} \alpha_i$. Then the group 
$Q/Q_I$ is isomorphic to a direct product of ${\bf Z}^{r - \# I}$ 
and a cyclic group of order $\gcd\{a_i|i \notin I\}$.

{\bf Proof}.
If $0 \notin I$, the lemma is obvious. If $0 \in I$, we may choose a basis
of $Q_I$ of the form $\alpha_0, \alpha_{i_1},  \alpha_{i_2}, \ldots$ where
$i_1, i_2, \ldots \geq 1$. After reordering of the set $\{1,\ldots, r\}$, the
matrix expressing this basis in terms of the basis of $Q$ takes the form
 \be
M \ =\ \left(
\begin{array}{c|c} 
a_1 \ldots a_s & a_{s+1} \ldots a_r \\ \hline
{\rm diag}(1,\ldots, 1) & 0 \\
\end{array}
\right)
  \ee
 It follows that all elementary divisors
of $M$ are $a:= \gcd\{a_{s+1}, \ldots a_r\}, 1,\ldots, 1$. Hence the maximal
finite subgroup of $Q/Q_I$ is a cyclic group of order $a$.

{\bf Remark 1}. Recall that the element $g \in G$ is called 
non--ad--exceptional
\cite{KT} if it can be written in the form $g = e^{2\pi i \beta}$, where
$(Ad\ g) x = x$, iff $[\beta, x] = 0$ ($\beta, x \in i \ Lie\ G$). It was
shown in Ref.\cite{KT} that $g$ is  non--ad--exceptional iff the set of 
integers $a_i$ for which the coordinates $s_i$ of $g$ are non--zero are 
relatively prime. It follows from our Theorem 1c that, for a 
non--ad--exceptional $g$, $\pi_1(G_g)$ is a free abelian group (or, 
equivalently,
the semi-simple part of $G_g$ is simply connected) and that for
a simply laced $G$ the converse holds as well. 

\vspace{.3cm}

\newpage

\begin{figure}
 \begin{center}
        \epsfxsize=150pt
        \epsfysize=0pt
        \vspace{-5mm}
        \parbox{\epsfxsize}{\epsffile{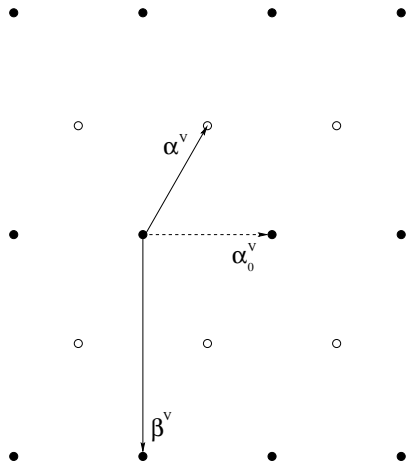}}
        \vspace{5mm}
    \end{center}
\caption{Coroot lattices for $G = G_2$ and its subgroup $\tilde G = 
[SU(2)]^2/{\bf Z}_2$. Blank circles mark out the nodes of $Q^\lor( G)$ not belonging
to $Q^\lor(\tilde G)$. }
\label{G2lat}
\end{figure}

 Fig. \ref{G2lat} provides an illustration for the Theorem 1
 in the simplest non--trivial
case of $G_2$. The lattice $Q^\lor( G = G_2)$ is formed by the coroots 
$\alpha^\lor$ and $\beta^\lor$ and the lattice $Q^\lor(\tilde G)$ for the
centralizer is formed 
by the (orthogonal) coroots $\beta^\lor$ and $\alpha_0^\lor = 2\alpha^\lor + 
\beta^\lor$. 
We see that  $Q^\lor( G)$ involves, indeed, extra nodes compared
to  $Q^\lor(\tilde G)$ (the blank circles in Fig. \ref{G2lat} ) and that
 $Q^\lor( G)/Q^\lor(\tilde G) \equiv \pi_1(\tilde G) = {\bf Z}_2$.
 (In this simple case,  $\pi_1(\tilde G)$
does not involve extra ${\bf Z}$ factors associated with the $U(1)$ factors
in the centralizer.)

\section{ Heisenberg pairs in compact Lie groups.}
\setcounter{equation}0

Given a central element $z$  in a compact Lie group $G$, a 
{\it Heisenberg pair with center} $z$ is a pair of elements $\tilde b, \tilde c
 \in G$ satisfying the commutation relation (\ref{Heis}). If in addition
$z$ has order $m$,  and one has
 \be
\label{bcm}
\tilde b^m \ = \ \tilde c^m \ = \ \left\{
\begin{array}{c}
1 \ \ \ \ \ \ {\rm if}\ m {\rm \ is \ odd} \\
z^{m/2} \ \ \ \ \ \  {\rm if}\ m {\rm\  is\  even}
\end{array}
 \right.
\ee
then $(\tilde b, \tilde c)$ is called a {\it standard} Heisenberg pair with 
center $z$. A Heisenberg pair $(\tilde b, \tilde c)$ is called {\it lowest
dimensional} if its orbit $G \cdot  (\tilde b, \tilde c)$ under conjugation
has minimal dimension (among Heisenberg pairs with center $z$).

{\bf Example 1}. Let $(\tilde b, \tilde c)$ be an irreducible Heisenberg pair
in $U(N)$ (i.e. there are no non--trivial subspace invariant with respect to
$\tilde b$ and $\tilde c$ ). Then $z = \epsilon I_N$ where $\epsilon$ is a 
primitive $m$-th root of 1. Let $v_1$ be an eigenvector for $\tilde b$:
$\tilde b v_1 \ =\ \beta v_1$, and let $v_{j+1} = \tilde c ^{j + 1} v_1$,
$j = 1, \ldots , N-1$. Since the pair $(\tilde b, \tilde c)$ is irreducible,
the vectors $v_j$ , $j = 1, \ldots , N$ form a basis of ${\bf C}^N$. Due to
(\ref{Heis}), we have
 \be
\label{bv}
\tilde b v_j \ = \ \epsilon^{j-1} \beta v_j,\ \ \ \  j = 1, \ldots, N \ .
 \ee
Since, due to (\ref{Heis}), $\tilde c$ permutes cyclically the eigenspaces of $\tilde b$, it follows
from irreducibility of the pair $(\tilde b, \tilde c)$ that all eigenspaces
of $\tilde b$ are 1 - dimensional. Hence, we have
\be
\label{cv}
\tilde c v_j &=& v_{j+1}, \ \ j = 1, \ldots , N-1 \nonumber \\
\tilde c v_N &=& \gamma v_{1} \ .
 \ee
Since $\tilde c^N v_1 $ is a multiple of $v_1$, we get by (\ref{Heis}) that 
$\epsilon^N = 1$. Therefore
 \be
\label{Neqm}
N \ = \ m  \ .
\ee
Thus, all irreducible Heisenberg pairs  $(\tilde b, \tilde c)$ in $U(N)$
with center $z = \epsilon I_N$, where $\epsilon$ is a primitive $m$-th root
of 1, are given up to conjugation by (\ref{bv}) --  (\ref{Neqm}), where $\beta$
and $\gamma$ are arbitrary constants of modulus 1.

If $(\tilde b, \tilde c)$ is a standard irreducible Heisenberg pair  in $U(m)$
with center $\epsilon I$, $\epsilon$ being a 
primitive $m$-th root of 1, then clearly 
 \be
\label{betgam}
\beta^m \ =\  \gamma \ =\ (-1)^{m+1} \ .
 \ee
It follows that there exists a unique up to conjugation such pair and it lies
in $SU(m)$. [For $m = 2$, the matrices (\ref{bv}), (\ref{cv}) with the 
condition
(\ref{betgam}) are reduced to $i\sigma_3$, $-i\sigma_2$ and, for $m=3,\ 
\epsilon = e^{2\pi i /3}$, they coincide with (\ref{HeisSU3}).]

Due to complete reducibility, we obtain that there exists  in $U(N)$ 
a Heisenberg pair
with center $z = \epsilon I_N$, $\epsilon$ being a primitive $m$-th root
of 1, iff $m$ divides $N$ and that any such pair is conjugate to a 
direct sum of $N/m$ such irreducible pairs. Moreover, in this case there
exists a unique up to conjugation standard Heisenberg pair, say $(\tilde b, 
\tilde c)$, which is a direct sum of irreducible such pairs, and  any 
Heisenberg pair in $U(N) $ or $SU(N)$ can be conjugated to a pair
 obtained by multiplying 
$(\tilde b, \tilde c)$ by a direct sum of matrices $\beta_i I_m$, 
$i = 1, \ldots, N/m$,  where $|\beta_i| = 1$ and, in the $SU(N)$ case,
$\prod_i \beta_i^m = 1$.

Note that the centralizer of the standard Heisenberg pair 
$(\tilde b, \tilde c)$ is isomorphic to the direct product of 
$\langle z \rangle$ and $SU(N/m)$  (here and further $\langle z \rangle$
denotes the cyclic group generated by $z$). It is also clear that this is
a  lowest dimensional Heisenberg pair with center
$\epsilon I_N$, where $\epsilon$ is a primitive $m$-th root
of 1.

{\bf Proposition 1.} Let $G$ be as in Theorem 1, let $z$ be a central 
element
of $G$ and let $(\tilde b, \tilde c)$ be a Heisenberg pair with center
 $z$. Let $T^0$
be a maximal torus in the centralizer of this pair in $G$. Then any 
Heisenberg
pair with center $z$ can be conjugated to a Heisenberg pair of the form  $(\tilde b t, \tilde 
c t')$, where $t,t' \in T^0$, and the rank of its centralizer 
always equals dim $T^0$.
Two pairs from ${\cal P}:= \tilde b T^0 \times \tilde c T^0$ can be
conjugated to each other iff they can be conjugated by an element
from the normalizer of ${\cal P}$ in $G$.

{\bf Proof}. Consider the group $G_{\tilde b, \langle z \rangle} \ =\ 
\{x \in G| x^{-1} \tilde b x \in \tilde b \langle z \rangle \}$. Its connected
 components
are $z' G_{\tilde b}$, where $z' \in \langle z \rangle$. The connected 
component 
containing $\tilde c$ is $z G_{\tilde b}$; it consists of all $\tilde 
c'$ such
 that $(\tilde b, \tilde c')$ is a Heisenberg pair with center $z$. The 
element $\tilde c$ acts on
 $G_{\tilde b, \langle z \rangle}$ by conjugation preserving the 
component $z G_{ \tilde b}$.
Recall that, by Gantmacher's theorem \cite{G}, 
{\it (i)} all conjugacy classes in a 
compact Lie group $K$
contained in a connected component of its element $g$ intersect the set
$g T^0$, where $T^0$ is a maximal torus in the centralizer of $g$
 in $K$, 
{\it (ii)} up to conjugacy, $T^0$ depends only on the connected component, {\it (iii)} two elements from $gT^0$ can be conjugated by an element of $K$ iff they can be conjugated by an element from the 
normalizer of $gT^0$ in $K$.
Hence, by the first part of Gantmacher's theorem, all conjugacy 
classes of the group $G_{\tilde b, 
\langle z \rangle}$ contained
in $z G_{\tilde b}$ intersect the set $\tilde c T^0$. 

Let $(\tilde b', \tilde c')$ be a Heisenberg pair  with center $z$. By above, 
we may assume
that $\tilde c' = \tilde c t', \ t' \in T^0$. But $\tilde b \in
 G_{\tilde c t', \langle z \rangle}$, and arguing as above we
see that $\tilde b'$ can be conjugated in the group $ G_{\tilde c t',\langle 
z \rangle}$ by the
conjugation action of $G_{\tilde c t'}$ into the set $\tilde b T^0$.
The fact that the rank
of the centralizer of $(\tilde b', \tilde c')$ equals dim $T^0$ follows
from the second part of  Gantmacher's theorem.
The last part of the proposition follows from the last part of Gantmacher's theorem.

\vspace{.3cm}

Now we will construct standard Heisenberg pairs $(\tilde b_m, \tilde c_m)$ in
 the 
universal covers of semi-simple parts of centralizers of exceptional elements.
This will provide us with a collection of commuting pairs $( b_m,  c_m)$
in the centralizers themselves.

Given a $G$--exceptional integer $m > 1$ , denote by $D_{m}$ the subdiagram
of $\hat D (G)$ consisting of the nodes $\alpha_j$ such that $a_j^\lor$ is
divisible by $m$, and let $\bar D_{m}$ be the complement of $ D_{m}$ in 
$\hat D(G)$. Denote by
$T_{m}'$ the subtorus of $T$ consisting of elements of the form 
 $ \exp \left \{ 2\pi i \sum_{j \in D_{m}} s_j \omega_j \right \}$
where $s_j \in {\bf R}$, and define the following 
(connected) subsets of $T_{m}'$:
  \be
 \label{subsets}
T_{m} \ &=&\ \left\{  \exp \left( 2\pi i \sum_{j \in D_{m}} s_j \omega_j
\right)  \right| \left. \sum_j a_j s_j = 0 \nonumber  \right\} \ , \\
T_{m}^{\rm alc} \ &=&\ \left\{ \exp \left( 2\pi i \sum_{j \in D_{m}} s_j 
\omega_j 
\right) \right| \left. \sum_j a_j s_j = 1, \ s_j \geq 0 \right \} \ .
 \ee
Note that $T_{m}^{\rm alc}$ is the intersection of $T_{m}$ with the
 fundamental alcove [consisting of the elements
 (\ref{excel}) satisfying
(\ref{conds})] and  $T_{m}$ is a subtorus in
$T_{m}'$ of codimension 1. Note also that dim $T_{m} = $ dim $
T_{m}^{\rm alc} = r_m$ 
, where $r_m$ is the number of nodes of $D_{m}$ minus 1.

It is clear that the centralizer of $T_{m}'$ in $G$ is
$T_{m} L_{m}$, where $ L_{m}$ is a semi-simple subgroup. Due to
Theorem 1b,c, $L_{m}$ is a connected semi-simple compact Lie group with Dynkin
diagram $\bar D_{m}$ and its fundamental group is a cyclic group of order
$m$. Let $\tilde L_{m}$ be the universal cover of $ L_{m}$, so that
$ L_{m} = \tilde L_{m}/{\bf Z}_{m}$, where ${\bf Z}_{m} \subset \tilde L_{m}$
is a cyclic group of order $m$. Let $z_m^{(1)}, \ldots , z_m^{(u_m)}$
be all elements of ${\bf Z}_{m}$ such that $\langle z_m^{(i)} \rangle = 
{\bf Z}_{m}$. Note that $u_m$ is the number of positive integers $\leq m$ 
relatively prime to $m$. A look at  Table \ref{Dynall} gives the 
following 
possibilities  for $\tilde L_{m}$  (provided that $m$ is $G$--exceptional), 
 for an $l$--laced $G$ :

\vspace{.5cm}

  \begin{tabular}{c|c|c|c|c|c} 
m&2&3&4&5&6 \\ \hline
$ \tilde L_{m}$& $[SU(2)]^{5-l}$ &  $[SU(3)]^{4-l}$ &  $[SU(4)]^2 \times 
SU(2)$
&  $[SU(5)]^{2}$  & SU(6) $\times SU(3) \times SU(2)$ \\ 
\end{tabular}

\vspace{.2cm}

\centerline{Table 3: The groups $\tilde L_{m}$.}

\vspace{.5cm}

It follows from Corollary 1 that ${\bf Z}_m$ is embedded diagonally in 
$\tilde L_{m}$. Since each $\tilde L_{m}$ is a product of special unitary
groups, using Example 1, one immediately constructs for each $z_m^{(i)},
\ i = 1, \ldots, u_m$, a standard Heisenberg pair $(\tilde b_m^{(i)},
\tilde c_m^{(i)})$ in $\tilde L_{m}$. Their images
$ b_m^{(i)}$ and $ c_m^{(i)}$ in $L_{m}$ form a commuting pair.

{\bf Theorem 2}. Let $G$ be as in Theorem 1 and let $m \geq 2$ be a 
$G$--exceptional integer. 

{\it (a)} Let $Q_{m}$ denote the connected semi-simple subgroup of $G$
obtained by removing only some ($\geq 1$ ) of the nodes $j$ from $\hat D(G)$
such that $a_j^\lor$ is divisible by $m$ and let $\tilde L_{m} \subset
\tilde Q_{m}$ be the inclusion corresponding to the inclusion of the Dynkin
diagrams. Then ${\bf Z}_{m}$ is central in $\tilde Q_{m}$ and $Q_{m} = 
\tilde Q_{m}/{\bf Z}_{m}$.

{\it (b)} Let $b,c \in Q_{m}$ be a commuting pair such that 
$\tilde b \tilde c
= z_m^{(i)} \tilde c \tilde b$ for (any) preimages of $b$ and $c$ in $Q_{m}$.
Then the pair $(b,c)$ can be conjugated by $ Q_{m}$ to $b_m^{(i)} T_{m}
\times c_m^{(i)} T_{m}$. The centralizer of all pairs from this subset
has rank $r_m$.

{\bf Proof}. {\it (a)} is established by a direct verification for maximal
$\tilde Q_{m}$'s. {\it (b)} follows from Proposition 1.

\section{Classification of commuting triples .}
\setcounter{equation}0

Denote by ${\cal T} \subset G \times G \times G$ the set of all commuting 
triples. 
For each $G$--exceptional integer $m \geq 2$, we construct the following $u_m$
families of commuting triples $(i = 1, \ldots, u_m)$:
  \be
 \label{FTTT}
F_m^{(i)} \ =\ T^{\rm alc}_{m} \times b_m^{(i)} T_{m} \times  c_m^{(i)} 
T_{m}\ \subset {\cal T}
 \ee
where $ b_m^{(i)},  c_m^{(i)} \in L_{m} \subset G$ are commuting pairs
constructed in Sect. 3.  It is clear from the construction that $T_{m}$ is
a maximal torus in the centralizer of each such triple.

Let $F_1^{(1)} \ =\ T^{\rm alc}_{1} \times T \times T$ and let 
  \be
 {\cal T}_m^{(i)} \ =\ G \cdot F_m^{(i)} \ \subset \ {\cal T}\ ,
 \ee
where $m$ is a $G$--exceptional integer and $i = 1, \ldots, u_m$. Note that
each $ {\cal T}_m^{(i)}$ is a connected component in the set of all commuting
triples in $G$. Since dim $F_m^{(i)} \ =\ 3r_m$ ($r_1 = r$) and the centralizer
in $G$ of a generic triple of $F_m^{(i)}$ is a torus of dimension $r_m$, we 
have
  \be
 \label{dimM}
{\rm dim}\   {\cal T}_m^{(i)} \ = \ {\rm dim} \ G + 2r_m \ .
  \ee
Triples from $\cup_i {\cal T}_m^{(i)}$ are called $m$--{\it exceptional}.

{\bf Theorem 3}. 
Let $G$ be the same as in Theorem 1. 

{\it (a)} The set ${\cal T}$ of commuting triples in $G$ is a disjoint union 
of the subsets $ {\cal T}_m^{(i)}$, where $m$ runs over all $G$--exceptional integers and $i= 1, \ldots, u_m$. 

{\it (b)} Rank of the centralizer of a triple from $ {\cal T}_m^{(i)}$ is
$r_m$.

{\it (c) } The sum over all components of ${\cal T}$ of the numbers $r_m + 1$
is equal to $h^\lor$.

{\bf Proof}. Let $(a,b,c)$ be a triple of commuting elements in $G$.
If $a$ is not exceptional, then by Theorem 1c, $\pi_1(G_a)$ is a free abelian
group, hence $G_a$ is a direct product of a torus $ T_1 \subset T$ and the
semi-simple part which is a connected  simply connected group. But then the
commuting pair $(b,c) \subset G_a$ can be conjugated to $T$ in $G_a$ (since the
centralizer of $b$ in $G_a$ is connected, the element $c$ can be conjugated
in $G_a \cap G_b$ to $T$). Thus, in this case $(a,b,c)$ is a trivial triple
(i.e. it can be conjugated to $T$). 

Let now $a$ be $m$--exceptional for $m \geq 2$. We may assume that
$a \in T^{\rm alc}_{m}$ (see Theorem 1a). Then the group $G_a$ is a product of one
of the groups $Q_{m}$ (see Theorem 2) and a torus, say $T_1$. Let
$\tilde G_a \ =\ T_1 \times \tilde Q _{m}$ so that
$G_a = \tilde G_a / {\bf Z}_{m}$ and let $\tilde b, \tilde c$ denote some preimages
of $(b,c) \in G_a$ in $\tilde G_a$. If $\tilde b \tilde c = \tilde c \tilde b$,
then the pair $(\tilde b, \tilde c)$ can be conjugated in $\tilde G_a$ to the
preimage of $T$ and the
triple $(a,b,c)$ is trivial. Finally, if $\tilde b \tilde c = z \tilde c 
\tilde b$ 
for a non--trivial element $z = z^{(i)}_d$ of ${\bf Z}_{m}$ of order $d$, where
$d$ divides $m$, then by Theorem 2, the pair $(b,c)$ can be conjugated in
$G_a$ to the family $\tilde b^{(i)}_d T_{d} \times 
\tilde c^{(i)}_d T_{d}$. It is also clear that, for all commuting triples
from a connected component, $z$ is the same and that $T_{d} 
\supset T_{m}\ ,\  T_{d}^{\rm alc} 
\supset T_{m}^{\rm alc}  $. This completes the proof of {\it (a)}.
{\it (b)} follows from Theorem 2.

Of course, one can check directly that in all cases one has
 \be 
 \sum_m (r_m + 1) u_m \ =\ h^\lor \ .
 \ee
Here is a unified proof which uses the classical fact that for any positive
integer $a$ one has
\be
  a \ =\ \sum_{j | a} u_j  \ . 
  \ee
Indeed, it follows from this formula applied to each $a_i^\lor$ that 
  \be 
h^\lor := \sum_{i=0}^r a_i^\lor \ =\ \sum_{j \geq 1} u_j N_j^\lor \ ,
 \ee
where $N_j^\lor = \# \{a_s^\lor | j \ {\rm divides}\  a_s^\lor
(s = 0, \ldots, r)\}$. On the other hand, $N_j^\lor = r_j + 1$.  

\vspace{.3cm}

The values of $r_m$ and $u_m$ for $m > 1$ are listed in Table \ref{rmum}.
The group $W_{m}$ in this table is the Weyl group of the torus 
$T_{m}$, i.e. the group
of linear transformations of $i Lie \ T_{m}$ generated by conjugations of
$G$ which leave $T_{m}$ invariant (  $T_{1}$ is just a 
maximal
torus and $W_{1}$ is the Weyl group $W$ of $G$).
We will exploit it in the following section
and in the Appendix B where  the calculation of $W_{m}$ for $m > 1$ is also 
explained (see Remark B2). The group $C_{m}$  is the semi-simple part of the 
centralizer 
of $T_{m}$ in $G$. It will be discussed and calculated below where we will
give an explicit construction of the moduli space for the commuting triples.
The same refers to the finite group $\Gamma_{m}$ in the last column of Table 4.

\setcounter{table}3 
  \begin{table}
\label{rmum}
 
\begin{tabular}{||l|c|c|c|c|c|c||} 
$ G$ & $m$ &$ r_m$ & $u_m$& $W_{m}$& $C_{m}$ &  $\Gamma_{m}$\\ \hline
$Spin(2r + 1)$ & 2 & $ r - 3$ & 1 & $ W_{B_{r-3}}$& $Spin(7)$& ${\bf Z}_2$ \\ \hline
$Spin(2r)$ & 2 & $ r - 4$ & 1 & $ W_{B_{r-4}}$ & $Spin(8)$ & ${\bf Z}_2$\\ \hline
$G_2$ & 2 & 0 & 1 & 1& $G_2$& 1 \\ \hline
$F_4$ & $\begin{array}{l} 2 \\3 \end{array}$ & $\begin{array}{c} 1 \\ 0 
\end{array} $ &  $\begin{array}{c} 1 \\ 2 \end{array}$& $\begin{array}{c} 
W_{A_1} \\ 1 \end{array} $ & $\begin{array}{c} Spin(7) \\ F_4 \end{array}$
& $\begin{array}{c} {\bf Z}_2 \\ 1 \end{array}$  \\ \hline
$E_6$ & $\begin{array}{l} 2 \\ 3 \end{array} $& $\begin{array}{c} 2 \\ 0 
\end{array} $
& $\begin{array}{c} 1 \\ 2 \end{array}$ & $\begin{array}{c}  
 W_{A_2} \\ 1 \end{array}$  & $\begin{array}{c} Spin(8) \\ E_6 \end{array}$ 
& $\begin{array}{c}  {\bf Z}_2 \times {\bf Z}_2  \\ 1 \end{array}$
  \\ \hline
$E_7$ &  $ \begin{array}{l} 2 \\ 3 \\ 4 \end{array} $
 & $ \begin{array}{c} 3 \\ 1 \\ 0 \end{array} $ & $\begin{array}{c} 1 \\ 2 \\ 
2 \end{array}$ &
$\begin{array}{c}  W_{C_3} \\  W_{A_1} \\ 1 \end{array} $&
$\begin{array}{c} 
Spin(8) \\ E_6 \\ E_7 \end{array}$ 
& $\begin{array}{c} 
{\bf Z}_2 \times {\bf Z}_2 \\ {\bf Z}_3 \\ 1 \end{array}$
\\ \hline
$E_8$ & $\begin{array}{l} 2 \\ 3 \\ 4 \\ 5 \\ 6 \end{array} $ & $
\begin{array}{c} 
4\\ 2 \\ 1\\ 0\\ 0 \end{array} $ & $\begin{array}{c}  1\\ 2 \\2 \\ 4 \\ 2 
\end{array} $ & $
\begin{array}{c}  W_{F_4} \\  W_{G_2} \\ W_{A_1} \\ 1 \\ 1 
\end{array} $ &
$\begin{array}{c} 
Spin(8) \\ E_6 \\ E_7 \\ E_8 \\ E_8 \end{array}$ 
& $\begin{array}{c} 
{\bf Z}_2 \times {\bf Z}_2 \\ {\bf Z}_3 \\ {\bf Z}_2 \\ 1 \\ 1 \end{array}$
 \\ \hline
\end{tabular}

\caption{non--trivial components of the moduli space, their Weyl groups, 
 the groups $C_{m}$ and  $\Gamma_{m}$.}
\end{table}

{\bf Remark 2}. Recall that, due to Theorem 1c, an element $g$ of $G$ can be
conjugated to $T^{\rm alc}_{m}$ iff the order of the maximal finite subgroup of
$\pi_1(G_g)$ is divisible by $m$. Pick an element $a_m \in T^{\rm alc}_{m}$.
Then the centralizer of any generic element $g \in a_m T_{m}$ is
$T_{m} L_{m}$, and therefore $g$ can be conjugated to $T^{\rm alc}_{m}$
by $W$. It follows that any element from $a_m T^{\rm alc}_{m}$
can be conjugated (by $W$) to  $T^{\rm alc}_{m}$. Hence we have
  \be
 E_m^{(i)}: \ =\ a_m T_{m} \times b_m^{(i)} T_{m} \times  c_m^{(i)} 
T_{m} \ \subset {\cal T}^{(i)}_m \ .
  \ee
 We proceed now to explicitly describe the moduli space 
 ${\cal M}_m^{(i)}$ of triples from ${\cal T}_m^{(i)}$ as a quotient of  
$E_m^{(i)}$ by an action of a finite group.

A triple from ${\cal T}$ is called {\it isolated} if its centralizer is finite.
It follows from Theorem 3 that the following are equivalent conditions for a
triple to be isolated:
\begin{itemize}
\item $r_m = 0$
\item ${\cal T}_m^{(i)}$ is a single $G$--orbit.
\end{itemize}
It is easy to see that the centralizer of an isolated triple $(a,b,c)$ 
is a product of three cyclic groups of order $m$ (generated by $a, b$ and 
$c$). A look at Table 4 provides a complete list of isolated triples
given by Table 5dvips triples
(for each $G$ and $m$ there are $u_m$ of them):

\vspace{.5cm}

\begin{tabular}{||c|c|c|c|c|c|c|c|c||} 
$ G$ & $Spin(7)$ & $Spin(8)$ & $G_2$ & $F_4$ & $E_6$ & $E_7$ & $E_8$ & $E_8$
  \\ \hline $m$ & 2& 2& 2& 3& 3& 4& 5& 6  \\ \hline
\end{tabular}

\vspace{.2cm}

\centerline{Table 5: Isolated triples.}

\vspace{.4cm}

Consider the centralizer $C'_{m}$ of $T_{m}$ and let $C_{m}$ denotes
the semi--simple part of $C'_{m}$. Due to Corollary 1, $C_{m}$ is a 
connected simply connected semi--simple subgroup of $G$. Note that 
$C_{m} \supset L_{m}$, hence \ rank$(C_{m}) \geq r - r_m$ , but since
$T_{m}$ commutes with $C_{m}$, we get also the reverse equality. Hence
 \be
{\rm rank}\  (C_{m}) \ =\  r - r_m\ 
  \ee
and $C'_{m} = C_{m}T_{m}$. It is clear that $E_m^{(i)} \subset
C'_{m} \times C'_{m} \times C'_{m}$ and therefore $ E_m^{(i)}
\cap (C_{m} \times C_{m} \times C_{m})$ is not empty. In other words,
$C_{m}$ contains an $m$--exceptional triple, say $(a,b,c)$. This triple
is isolated in $C_{m}$, since in the contrary case the rank of its 
centralizer in $G$ would be greater than $r_m$, in contradiction with Theorem
3b. 

Thus $C_{m}$ is a semi-simple subgroup of $G$ containing an isolated 
$m$--exceptional triple
, and its Dynkin diagram is a subdiagram of $D(G)$
which consists of $r - r_m$ nodes. A look at Table 5 establishes
that the only possibilities for $C_{m}$ are those listed in Table \ref{rmum}.

{\bf Remark 3}. We have obtained a slightly more canonical construction of the
sets $E_m^{(i)},\ i = 1, \ldots, u_m$. Let $k_1, \ldots, k_{u_m}$ be all 
integers between 1 and $m$ relatively prime to $m$. Pick an isolated triple
$(a_m, b_m, c_m)$ in the subgroup $C_{m}$. Then
 \be
E_m^{(i)} \ =\ a_m T_{m} \times b_m T_{m} \times  c_m^{k_i} T_{m}
 \ee

Denote by $N_{m}$ the normalizer of the set $E_m^{(i)}$ in $G$ (clearly, it
is independent of $i$), i.e. $N_m = 
\left\{g \in G | g E_m^{(i)} g^{-1} \ \subset E_m^{(i)} \right\}$. 
Note that the
centralizer of $E_m^{(i)}$ in $G$ is $T_{m}$. Let $R_{m} = 
N_{m}/T_{m}$; then the moduli space ${\cal M}_m^{(i)}$ of triples from
${\cal T}_m^{(i)}$ is the quotient
$${\cal M}_m^{(i)} \ =\ E_m^{(i)}/R_{m} \ ,$$
where $R_{m}$ is a finite group that we are about to compute. 
This follows from the last part of Proposition 1. For 
example, in the case of trivial triples one has: 
${\cal M}_1^{(1)} \ =\ (T \times T \times T)/W$ (with the diagonal action
of $W$). 

Let $N (T_{m} ) \  = \ 
\left\{g \in G | g T_{m} g^{-1} \ \subset T_{m} \right\}$. Recall that
$W_{m} = N (T_{m} )/ (C_{m} T_{m}) $ is the Weyl group of 
$ T_{m}$. One, clearly, has
  \be
 N_m = 
\left\{g \in N(T_{m}) | g a_m g^{-1} \ \in a_m T_{m},\ 
 g b_m g^{-1} \ \in b_m T_{m}, \ 
 g c_m g^{-1} \ \in c_m T_{m} \right\} \ .
 \ee
Since there always exists a triple in $E_m^{(i)}$ for which the Weyl group
of the centralizer is $W_{m}$ (cf. Table 7), we obtain:
  \be
R_{m} \ =\ W_{m} A_{m}\ , 
 \ee
where $W_{m}$ normalizes $A_{m}$ and
  \be
A_{m} \ =\ \frac{
\left\{ g \in C_{m} | g (a_m, b_m, c_m) g^{-1} \ =\ 
(a_m t_1, b_m t_2, c_m t_3) \ \ {\rm for\ some}\ \ t_1, t_2, t_3 \ \in T_{m}
\right\}}
{\left\{ {\rm centralizer\ of}\ (a_m, b_m, c_m)\ {\rm in} \ C_{m}\right\}}\ .
   \ee
But $t_1, t_2, t_3$ lie also in $C_{m}$, hence they are central in $C_{m}$.
Note that multiplying each element of the triple $(a_m, b_m, c_m)$ by a central
element of $C_{m}$ again gives an isolated triple of $C_{m}$. Hence we 
have:
  \be
  A_{m} \ =\ \Gamma_{m} \times \Gamma_{m} \times \Gamma_{m} \ ,
  \ee
where $\Gamma_{m} \ =\ {\rm Center} (C_{m} ) \cap T_{m}$. 
We thus have proven the following 
theorem:

{\bf Theorem 4.} The connected component ${\cal M}_m^{(i)}$ of the moduli
space of $m$--exceptional triples in $G$ looks as follows:
 \be
\label{moduli}
{\cal M}_m^{(i)} \ =\ \left. \left( 
\frac {a_m T_{m}}{\Gamma_{m}} \times \frac {b_m T_{m}}{\Gamma_{m}}
\times \frac {c_m^{k_i} T_{m}}{\Gamma_{m}} \right) \right/ W_{m}
  \ee
Here $(a_m, b_m, c_m)$ is an isolated $m$--exceptional triple in $C_{m}$,
$k_i$ are all integers between 1 and $m$ relatively prime to $m$, 
$\Gamma_{m} \ =\ {\rm Center} (C_{m} ) \cap T_{m}$ acting by multiplication,
and $W_{m}$ is the Weyl group of $T_{m}$ acting diagonally. The groups $W_m$ 
are listed
in the corresponding column of Table 4.
(They happen to coincide with the Weyl group of the
centralizer of a lowest-dimensional $m$--exceptional triple; see Remark B2 and
Corollary B2.)

\vspace{.3cm}

The only remaining question is {\it what} are the groups $\Gamma_{m}$. We have
resolved it by explicit calculations described in Appendix C. The results
are given in the last column of Table \ref{rmum}. It turns out that in almost
all cases $\Gamma_{m}$ coincides with the center of  $C_{m}$. The only 
exception
are the groups $Spin(2r)$, $r > 4$; the center of $C_{2} = Spin(8)$ in this
 case is
${\bf Z}_2 \times {\bf Z}_2$, but the group  $\Gamma_{2}$ is just ${\bf Z}_2$.

{\bf Remark 4}. Let $G$ be as in Theorem 1 and let $\nu$ be a diagram 
automorphism of $G$ of order $k = 2$ (for types $A, D, E_6$) or 3 (for type
$D_4$). Let $G'$ be a connected component of the semidirect product of 
 $\langle \nu \rangle$
and $ G$, containing $\nu$. Then Theorems 1-4 hold if instead of $G$ one takes
$G'$ and instead of the extended
Dynkin diagram (denoted by $X_r^{(1)}$ in \cite{K}), one takes the ''twisted
diagrams`` $X_r^{(k)},\ \ k = 2,3$. In particular, the sum $\sum_m (r_m + 1)
 u_m$  equals the dual Coxeter number  of $X_r^{(k)}$ which is equal to the
Coxeter number  of the group $G$.

\section{Counting quantum vacua. Examples.}
\setcounter{equation}0

Consider a connected component ${\cal M}_m^{(i)}$ in the moduli space of all 
commuting triples. As we have seen, it presents a subset of the product of
three (shifted) tori $T^{r_m}$ of dimension $r_m$ (or rather a quotient of this
product  by action of a finite group), 
 where $r_1$ is the rank of the group and  $r_m$ for all non--trivial 
components for all groups are listed in Table 
\ref{rmum}.
In the full analogy with (\ref{vacT}), ${\cal M}_m^{(i)}$ gives rise to 
$r_m + 1$
quantum vacuum states 
  \be
\label{vacTsmall}
\Psi \ \sim\  1, \sum_{i}^{r_t}  \epsilon^{\alpha\beta} 
\lambda_\alpha^i \lambda_\beta^i, \ldots , \left( \sum_{i}^{r_m} 
\epsilon^{\alpha\beta} 
\lambda_\alpha^i \lambda_\beta^i \right)^{r_m}\ ,
\ee
where $i$ mark the Cartan generators forming an orthonormal basis on
 the small torus
$T^{r_m}$ and $\sum_{i}^{r_m} q^i q^i$ is the quadratic invariant
of the Weyl group $W_{m}$ of  $T^{r_m}$ (see Table \ref{rmum}, last column).
  \footnote{Choosing a particular embedding $T^{r_m} \subset G$ amounts to 
fixing a particular gauge. After that, one should still require the invariance
with the respect to a discrete group of Weyl transformations which is a 
remnant of the original gauge group after the embedding is fixed. The relevant
groups $W_{m}$ are listed in Table 4.} 
Note that  the powers higher than $r_m$ in Eq. (\ref{vacTsmall}) would give 
just zero due to anticommuting nature of $\lambda_\alpha^i$. The Appendix B
is devoted to the proof of the fact that {\it all} 
linear independent elements of the Grassmann algebra with the base
$\lambda_\alpha^i, \  \alpha = 1,2$,  invariant under the action of the Weyl
group are reduced to the powers of quadratic invariant which means that 
all quantum vacuum states associated with the component ${\cal M}_k$ 
are, indeed, listed in Eq.(\ref{vacTsmall}). 
We have all together
  \be
\label{quall}  
 {\rm Tr} (-1)^F \ =\ \sum_k (r_{m} + 1)_k 
  \ee
where the sum runs over all components ${\cal M}_k$.
\footnote{The wave functions in Eq.(\ref{vacTsmall}) depend only on the 
elements of the algebra and do not know about the global structure of the 
triples' moduli space studied carefully in the previous section. Therefore,
this structure does not affect the counting (\ref{quall}).}
 From Table \ref{rmum},
 we have
$(1 + 1) \cdot 1 + (0 + 1) \cdot 2 = 4$ extra states (associated with the 
components with $m > 1$) for $F_4$, $3 \cdot 1 + 1 \cdot 2 = 5$ extra states
for $E_6$, $4\cdot 1 + 2 \cdot 2 + 1 \cdot 2  = 10$ extra states for
$E_7$, and $5\cdot 1 + 3 \cdot 2 + 2 \cdot 2 + 1 \cdot 4 + 1 \cdot 2 = 21$
extra states for $E_8$. As is seen from Table 1, adding these states to
$r_{G} + 1$ gives exactly $h^\lor$ in all cases.

 This latter fact has been established by Theorem 3c. Another way
to understand it is the following. We have seen that the extra components
${\cal M}_k$ are associated with the nodes with $a_i^\lor > 1$ on the Dynkin
diagrams. Just by construction, the sum (\ref{quall}) over the components
with $m > 1$ coincides with the
sum $\sum_j (a_j^\lor - 1)$ over all such nodes. Consider, e.g. the node
${\bf 6}$ with $a_{\bf 6}^\lor = 6$ on the Dynkin diagram for $E_8$.  
It gives rise to
two different isolated triples with $m = 6$ (and, correspondingly, to two 
different  quantum states).  Together with two nodes with $a_i^\lor = 3$,
it gives rize to the component ${\cal M}_k$ based on the torus $T^{r_m = 2}$ 
( and if the node ${\bf 6}$ were not present
in the superposition, the dimension of the torus would be one unit less, and
we would have not 3, but just 2 vacuum states ).  Finally, 
together with the nodes with  $a_j^\lor = 2,4$, it forms the torus $T^4$. As we
have here $m=3$, there are two different components ${\cal M}_k$ involving
such torus and, correspondingly, two extra vacuum states brought about by the
node  ${\bf 6}$. All together, this node gives rise to $2 + 1 + 2 = 5 =
a_{\bf 6}^\lor - 1$ extra states. The same counting works for all
 other nodes.

But the sum $\sum_j (a_j^\lor - 1)$ is just the difference $h^\lor - 
r_{G} - 1$.
Indeed, $h^\lor$ coincides with the sum of all labels $a_i^\lor$ in the
 extended
Dynkin diagram \cite{K} while $r_{G} + 1$ is just the number of its nodes.

At the end, we list again our results for higher exceptional groups. Further
explicit formulae can be found in Appendix C.

\begin{itemize}
\item $F_4$. {\it i)} There
is a root ${\bf 3}$ with $a_{\bf 3}^\lor = 3$. According to Theorem 1, 
there is 
only one associated element $g = \exp\{(2\pi i/3)\ \omega_{\bf 3}\}$.
The centralizer of this element is
 $SU(3) \times SU(3)$ (the Dynkin diagram for the centralizer is obtained from
the extended Dynkin diagram of $F_4$ by removing the node
${\bf 3}$) factorized over the common center ${\bf Z}_3$.
\footnote{Note that the element $\tilde g = g^2$ has also this property, but
it gives nothing new because it is equivalent to $g$ by conjugation. This is
guaranteed by Theorem 1a.} ${\bf Z}_3$ has two non--trivial elements and, 
correspondingly,
there are {\it two} different non--trivial Heisenberg pairs in $SU(3)$ : the 
pair in Eq. (\ref{HeisSU3}) and its permutation. We have proven earlier [see
the remark after Eq.(\ref{betgam})]  that 
 all other Heisenberg  pairs can be reduced to (\ref {HeisSU3}) by 
conjugation. From this, we have two different isolated triples $\{\Omega_i\}$
and two different corresponding quantum vacuum states. Note that, in this case,
the second triple is obtained from the first one just by reordering
$\Omega_2 \leftrightarrow \Omega_3$, but they are still two different triples
from our viewpoint. They describe two {\it different} flat connections on
$T^3$ which cannot reduced one to another by gauge transformation.

{\it ii)} There are two nodes (the long root ${\bf 2}$ and the short
root ${\bf 4}$) with  $a_i^\lor = 2$. The 
centralizer of $g = \exp \{2\pi i (s_{\bf 2} \omega_{\bf 2} + s_{\bf 4} 
\omega_{\bf 4})\},\ \ 2s_{\bf 2} +  4s_{\bf 4} = 1$, is non--trivial. If
$s_{\bf 2}, s_{\bf 4} \neq 0$, it is $[SU(2)]^3/{\bf Z}_2 \times U(1)$. 
Correspondingly, there is
a family $ T^1 \times T^1 \times T^1$ [up to a factorization over a certain 
discrete group --- see Eq.(\ref{moduli}) and also Appendix C for more 
details ] of non--trivial commuting triples which gives rise to two extra 
quantum vacua.

\item $E_6$.  
{\it i)} The element
$\exp\{(2\pi i/3)\ \omega_{\bf 3}\}$ associated with the node  with
$a_i = 3$
 \footnote{All the roots for the groups $E_{6,7,8}$ are long and there is no 
distinction
between $a_i$ and $a_i^\lor$.}
has a centralizer  $[SU(3)]^3/{\bf Z}_3$. It gives two different isolated triples
and two corresponding vacua.

{\it ii)} Three nodes with $a_i = 2$ produce a family
$\subset T^2 \times T^2 \times T^2$ of non--trivial commuting triples. There are
3 quantum vacuum states associated with that.

\item $E_7$. 
{\it i)} The centralizer of the  element
$\exp\{\pi i/2\  \omega_{\bf 4}\}$ associated with the node  with
$a_i = 4$ is $G_{\bf 4} = \tilde  G_{\bf 4}/{\bf Z}_4$ where 
$\tilde  G_{\bf 4} = [SU(4)]^2 \times SU(2)$ and the factorization over 
${\bf Z}_4$ implies the following identification of the elements of the center
of $\tilde  G_{\bf 4}$: 
  \be
1 \ \equiv \ (\pm i I_4, \ \pm i I_4,\  -I_2) \ \equiv \ (- I_4, \ - I_4,\ 
 I_2)\ .
  \ee
 The elements $(\pm i I_4, \pm i I_4, -I_2)$ are of the fourth order. For
each such element, there is unique up to conjugation Heisenberg pair is
$\tilde G_{\bf 4}$: $(\tilde b^{(4)}_\pm, \tilde b^{(4)}_\pm, i\sigma_3)$ and
$(\tilde c^{(4)}_\pm, \tilde c^{(4)}_\pm, i\sigma_2)$ , where $( \tilde 
b^{(4)}_\pm,   \tilde c^{(4)}_\pm )$ are the standard Heisenberg pairs with
center $\pm I_4$ in $SU(4)$. The corresponding elements in the centralizer
commute. Thereby, we have two different isolated triples and two new vacua.

{\it ii)} 
The centralizer $G_{\bf 4}$
 admits also Heisenberg pairs based on the 2--order element of ${\bf Z}_4$. A
 standard Heisenberg pair in $\tilde G_{\bf 4}$
associated with that has the form $(\tilde b^{(2)}, \tilde b^{(2)}, 1)$; 
$(\tilde c^{(2)}, \tilde c^{(2)}, 1)$, where $(\tilde b^{(2)}, \tilde c^{(2)})$
is the standard Heisenberg pair with center $-I_4$ in $SU(4)$. 
The centralizer of the triple formed by a canonical exceptional element
$\exp\{i \pi/2 \ \omega_{\bf 4} \}$ and the commuting elements in $G_{\bf 4}$
corresponding to this Heisenberg pair is $[SU(2)]^3$ where first two
$SU(2)$ factors are the centralizers of the pair $(\tilde b^{(2)}, 
\tilde c^{(2)})$ in each $SU(4)$ and the third one is the whole $SU(2)$ factor
in  $\tilde G_{\bf 4}$. 
 The 3 + 3 -- parametric family of triples following from that is
actually a subset of a larger family  lying in $T^3 \times T^3 \times T^3$  
which is based on a generic
exceptional element formed by 
the node ${\bf 4}$ together with three other nodes  with
$a_i = 2$. The centralizer of a generic such element is
   $[SU(2)]^4/{\bf Z}_2 \times [U(1)]^3$ and the centralizer of a generic such
triple is $[U(1)]^3$. From this, we obtain 4 extra quantum vacua.

{\it iii)} There is a one-parametric family of exceptional elements associated
with the nodes  with $a_i = 3$.  The 
centralizer of a generic such element is $[SU(3)]^3/{\bf Z}_3 \times U(1)$.
 Correspondingly, there are two 
different 
components $\subset T^1 \times T^1 \times T^1$ in the moduli space of the 
triples ${\cal M}$, each of them giving 2 extra vacua.

\item $E_8$. 
{\it i)} The element
$\exp\{\pi i/3\ \omega_{\bf 6}\}$ associated with the node  with
$a_i = 6$ has the centralizer $G_{\bf 6} = \tilde  G_{\bf 6}/{\bf Z}_6$ where 
$ \tilde G_{\bf 6} \ =\ SU(6) \times SU(3) \times SU(2)$ and the factorization
over ${\bf Z}_6$ implies
  \be
1 \ \equiv \ (e^{ \pi i/3 } I_6, \  e^{2 \pi i/3 } I_3,\  -I_2) \ \equiv \ 
\  (e^{2 \pi i/3 } I_6,\  e^{4 \pi i/3 } I_3,\ I_2)\ \equiv \ \cdots 
  \ee
The center of $\tilde G_{\bf 6}$ 
 has two elements of the
6-th order which are identified to 1 after factorization. They give rise
to    two different  Heisenberg pairs in $\tilde G_{\bf 6}$ which 
 correspond to commuting elements in $G_{\bf 6}$. We have 
two  isolated triples and two extra vacua.
Besides, ${\bf Z}_6$ has 2 elements of order 3 and an element of order 2. As was 
also the case for $E_7$, Heisenberg pairs associated with that are ''absorbed``
in the families {\it iii)} and {\it iv)} below, formed by a combination of
several nodes.

{\it ii)} The element
$\exp\{2\pi i/5\ \omega_{\bf 5}\}$ associated with the node  with
$a_i = 5$ has the centralizer  $[SU(5)]^2/{\bf Z}_5$. There are four
 different  Heisenberg pairs in $SU(5)$ with the property
$bc = \epsilon cb$, $bc = \epsilon^2 cb$, $bc = \epsilon^3 cb$ and 
$bc = \epsilon^4 cb$,
$\epsilon = \exp\{2\pi i/5\}$. Correspondingly: 4 isolated triples and 4 extra
vacua.

{\it iii)} There are two distinct components $\subset T^1 \times T^1 \times 
T^1$ associated with a 1--parametric mixture of the 4--exceptional elements
$\exp \{\pi i/2  \ \omega_{\bf 4}\}$ and   $\exp \{\pi i/2 \  
\omega_{\bf 4'}\}$ associated with two different nodes 
 with $a_i = 4$.  Each such component gives 2 vacua.

{\it iv)} There are two distinct components $\subset T^2 \times T^2 \times 
T^2$ formed by
the nodes ${\bf 6}$ and two nodes ${\bf 3}$, each of them giving 3 vacua.

{\it v)} The component  lying in $T^4 \times T^4 \times T^4$ formed by the 
nodes ${\bf 6}$, two nodes ${\bf 4}$ and two nodes ${\bf 2}$ gives 5 extra vacua.

\end{itemize}

{\bf  Note.}
A few days before we sent the first version of this paper to the archive,
another paper on this subject appeared
  \cite{Keur} where an explicit construction 
of the
set of non--trivial triples with $m=2$ for all exceptional groups has been done.
It was followed by the paper \cite{Keur2} where the structure of triples with 
 $m = 3,4,5,6$ was analyzed.  The results  agree with
ours. The question of completeness of this classification was not studied,
however, in these papers. The theorems proven here guarantee that, indeed,
{\it all} non--trivial triples lie in one of the families discussed above and
the total number of quantum states always coincides, indeed, with the dual
Coxeter number of the group.

{\bf Acknowledgements:} V. K. wishes to thank E.B. Vinberg for  very useful 
correspondence and remarks. A. S. is indebted to V. Rubtsov for useful 
discussions.

\section*{Appendix A: Heisenberg pairs in all connected simply connected
compact Lie groups.}
\setcounter{equation}0
\renewcommand{\theequation}{A.\arabic{equation}}

Here is a general way to construct a  Heisenberg pair $(\tilde b, \tilde
c)$ with center $z$ in a simply connected simple compact Lie group $G$
(for semi-simple $G$ we just take the product of the $(\tilde b_i, \tilde
c_i)$ over all simple factors $G_i$). 

Recall that all non--trivial central 
elements of $G$ are of the form $e^{2\pi i \omega_j}$ where $j$
is such that $a_j = 1$. Pick such a $j$ and consider the set of roots 
$\{\alpha_0, \alpha_1, \ldots, \alpha_r\} \backslash \alpha_j$. This is 
another set of
simple roots, hence there exists a unique element $w_j$ of the Weyl group
$W$ of $G$ which transforms it to the set ${ \alpha_1, \ldots, \alpha_r}$, 
hence $w_j \alpha_j = \alpha_0$. This gives us a canonical embedding of the
center of $G$ in $W$:  $e^{2\pi  i \omega_j} \to w_j$, and, at the same time,
in the group of symmetries of the extended Dynkin diagram $\hat D(G)$.

Let $z = e^{2\pi i \omega_j}$ be a central element of $G$ of order $ m \geq 2$.
Consider the corresponding element of the Weyl group $w_j$. and denote by
 $\langle w_j \rangle $ a 
Let $I_{\langle w_j \rangle} = \langle w_j \rangle \cdot \alpha_0$ 
(the notation $\langle w_j \rangle$ stands like before for  the 
cyclic subgroup of $W$ generated by the element $w_j$)
be the subset of the set of nodes
of $\hat D (G)$ obtained by the action of the group
  $\langle w_j \rangle $
on the node $\alpha_0$ ($a_s = 1$ for all nodes from
 $I_{\langle w_j \rangle}$).

Let 
\be
\label{tildeb}
\tilde b = \sigma_{\vec{s}}
 \ee
 as in Eq. (\ref{excel}), where $\vec{s}$
is defined by $s_i = 1/m$ if $i \in I_{\langle w_j \rangle}$ 
and $s_i = 0$ otherwise. Write $w_j
= r_{\gamma_1} \cdots r_{\gamma_k}$ as a shortest product of reflections with 
respect to some (not necessarily simple) roots 
$\gamma_1, \ldots, \gamma_k$ and let
 \be
\label{tildec}
\tilde c \ =\ q_{\gamma_1} \cdots  q_{\gamma_k} \ ,
 \ee
where $q_{\gamma_t}$ is the Pauli matrix $ i \sigma_2$
in the $SU(2)$ corresponding to the  root $\gamma_{t}$
[or in other words, $q_{\gamma_t} = \exp\left\{ \frac \pi 2 \left(T_{\gamma_t} 
- T_{-\gamma_t} \right) \right \}$].
The conjugation of any element $h$ of the Cartan subalgebra ${\mathfrak h}$ 
with the group
element  $q_{\gamma_t}$ presents the reflection with respect to the root
$\gamma_t$:
\be
\label{refl}
 (q_{\gamma_t})^{-1} h  q_{\gamma_t} \ =\ h - \frac{2 \langle h, 
\gamma_t^\lor \rangle}{\langle \gamma_t^\lor, 
\gamma_t^\lor \rangle} \gamma_t^\lor
\ee
Thereby, Eq.(\ref{tildec}) presents a particular lifting of $w_j$ in $G$ :
for the elements of the maximal torus,  $\tilde c^{-1} 
e^{i h} \tilde c \ = 
\ e^{i  w_j h}$, and the action of this conjugation can of course be also
defined for all other elements of $G$. The choice (\ref{tildec}) is convenient,
but one could take 
for $\tilde c$ any other such lifting.

{\bf Theorem  A1.} Let $G$ be as in Theorem 1 and let 
$z = e^{-2\pi i \omega_j}$ be a central element of $G$ of order $ m \geq 2$.

{\it (a)} The pair $(\tilde b, \tilde c)$ given by  (\ref{tildeb}) and
(\ref{tildec}) is a Heisenberg pair with center $z$.

{\it (b)} Any Heisenberg pair with center $z$ in $G$ can be conjugated to a 
Heisenberg pair of the form $(\tilde b t, \tilde c t')$, where  $(\tilde b, 
\tilde c)$ is given by  (\ref{tildeb}) and (\ref{tildec}) and $t,t'$ lie in
a fixed maximal torus $T^0$ of the centralizer of $(\tilde b, \tilde c)$.
Two pairs from ${\cal P}:= (\tilde b T^0, \tilde c T^0)$ can be
conjugated to each other iff they are conjugated by the normalizer 
of ${\cal P}$.

{\it (c)} The centralizer of a lowest dimensional Heisenberg pair of $G$
with center $z$ is the 
 direct product of $\langle z \rangle$ and a connected subgroup
$K$ listed in Table 6.

\setcounter{table}5
\begin{table}
\label{Heiscent}

\begin{tabular}{||l|c|l||} 
$ G$ & $m$ &$ K$ \\ \hline
$SU(N)$ & $m | N$ &$ SU(N/m)$ \\ \hline
$Sp(2N)$, $N$ odd  & 2 & $SO(N)$ \\ \hline
$Sp(2N)$, $N$ even  & 2 & $Sp(N)$ \\ \hline
$Spin(N)$, $N$ odd & 2 & $Spin(N-2)$ \\ \hline
$Spin(N)$, $N$ even & 2 & $Spin(N-3)$ \\ \hline  
 $Spin(N)$, $N$ even, $4 /\!\!| N$ & 4 & $Spin(N/2-2)$ \\ \hline  
$Spin(N)$, $N$ even, $4 | N$ & $2'$ & $Sp(N/2)$ \\ \hline 
$E_6$ & 3 & $G_2$ \\ \hline
$E_7$ & 2 & $F_4$ \\ \hline
\end{tabular}

\caption{Centralizers of lowest dimensional Heisenberg pairs.
Case $m=2$ for $Spin(N)$ stands for $z = -1$ and  case $m = 2'$ for
$Spin(4p)$ stands for $z = \gamma_{4p + 1} = \prod_{i=1}^{4p} \gamma_i$.}

\end{table}

{\bf Proof}. It is easy to see that $(\tilde b, \tilde c)$ is a Heisenberg
pair with center $z$. Indeed,  $\tilde c^{-1} \tilde b \tilde c \ = 
\ e^{i w_j \lambda}$, where $\lambda \ = (2 \pi /m) \sum_{s \in 
I_{\langle w_j \rangle} \backslash 
0} \omega_s$ and one can check that $e^{i (w_j \lambda - \lambda)} = z$, 
hence 
 $ \tilde b \tilde c \ =\ z  \tilde c \tilde b $ 
\footnote{Let us illustrate how this general construction works in the simplest
non--trivial case of $G = SU(3)$. Fundamental coweights are
$\omega_1 = (2\alpha_1 + \alpha_2)/3 = (1/3) {\rm diag} (2,-1,-1)$, \ 
$\omega_2 = (2\alpha_2 + \alpha_1)/3 = (1/3) {\rm diag} (1,1,-2)$. We have
$\exp\{-2\pi i \omega_1\} =  \epsilon  {\bf 1}$ and 
$\exp\{-2\pi i \omega_2\} =  \epsilon^2 {\bf 1}$\ 
($\epsilon = \exp\{2\pi i/3\}$). The corresponding elements of 
the Weyl group present the cyclic permutations of the eigenvalues of the 
diagonal $SU(3)$ matrices: $w_1:\ (231)$ and $w_2:\ (312)$. 
In this case, $\langle w_1 \rangle \ =\ \langle w_2 \rangle $,
 $I_{\langle w_j \rangle}$ coincides with the whole extended Dynkin diagram,
 and
$\tilde b = \exp \{2\pi i (\omega_1 + \omega_2)/3\} \ = {\rm diag} (\epsilon,
1, \epsilon^2)$.
Pick up $w_1 = r_{\alpha_2} r_{\alpha_1}$. Eq.(\ref{tildec}) is reduced to
$$
\tilde c \ =\  \left( \begin{array}{ccc}
1& 0&0 \\ 0 & 0& 1 \\ 0&-1& 0 \end{array} \right)
\left( \begin{array}{ccc}
0& 1&0 \\ -1 & 0& 0 \\ 0&0& 1 \end{array} \right) \ =\ 
\left( \begin{array}{ccc}
0& 1&0 \\ 0 & 0& 1 \\ 1& 0 & 0 \end{array} \right)\ . $$
We have $\tilde b \tilde c = \epsilon \tilde c \tilde b$ and the pair $(\tilde
b, \tilde c)$ thus constructed is equivalent by conjugation to the pair
(\ref{HeisSU3}).}.
{\it (b)} follows 
from Proposition 1. Finally, the centralizer of $\tilde b$ in $G$ is obtained
by removing all nodes $j$ with $a_j = 1$ from $\hat D(G)$. 
Multiplying, if
necessary, $\tilde c$ by  an element from a maximal torus from the centralizer
of $(\tilde b, \tilde c)$, we may assume that $\tilde c$ induces a diagram
automorphism of $G_{\tilde b}$ for which the fixed point sets are well--known
(see \cite{K}, Chapter 8). This proves {\it (c)}.
\footnote{{\it Example}: Consider the group $E_7$. The center of  $E_7$
is ${\bf Z}_2$ and the extended Dynkin diagram of  $E_7$ has  ${\bf Z}_2$ mirror symmetry.
The centralizer of $\tilde{b}$ in $E_7$ is $E_6$. The centralizer of the
pair $(\tilde{b}, \tilde{c})$ is a subgroup of $E_6$ invariant under the
conjugation corresponding this mirror symmetry. The Dynkin diagram for
such a subgroup is obtained from the Dynkin diagram of $E_6$ by gluing together
two pairs of the (co)roots. This is the Dynkin diagram of $F_4$ [the embedding
$F_4 \subset E_6$ thus constructed
will be discussed in some more details in Appendix C: see Eq.(\ref{F4E6})]. 
That
is how the result quoted in the last line of the last column of Table 6 is
obtained, and all other cases can be treated in a similar way. To avoid 
confusion, note that, in contrast to the centralizer of a single element, 
the centralizer of a Heisenberg pair need not, and generally is not a regular
subgroup of $G$.}

{\bf Remark A1}. Theorem A1 gives a classification of all up to conjugacy
commuting pairs in connected compact (not necessarily simply connected)
Lie groups $G$. In particular, this theorem shows that connected components
of the set of pairs of commuting elements of $G$ are in one-to-one 
correspondence with elements of the maximal finite subgroup of $\pi_1(G)$.

Using Table 6, one immediately gets the list of centralizers of
 $m$--exceptional triples $(a,b,c)$,
 where $a = \exp\{ 2\pi i \omega_j / a_j \},\ 
m = a_j^\lor$ and $(b, c)$ is the lowest dimensional pair in $G_a$. 

\begin{table}
\label{centtri}

\begin{center}
        \epsfxsize=400pt
        \epsfysize=0pt
        \vspace{-5mm}
        \parbox{\epsfxsize}{\epsffile{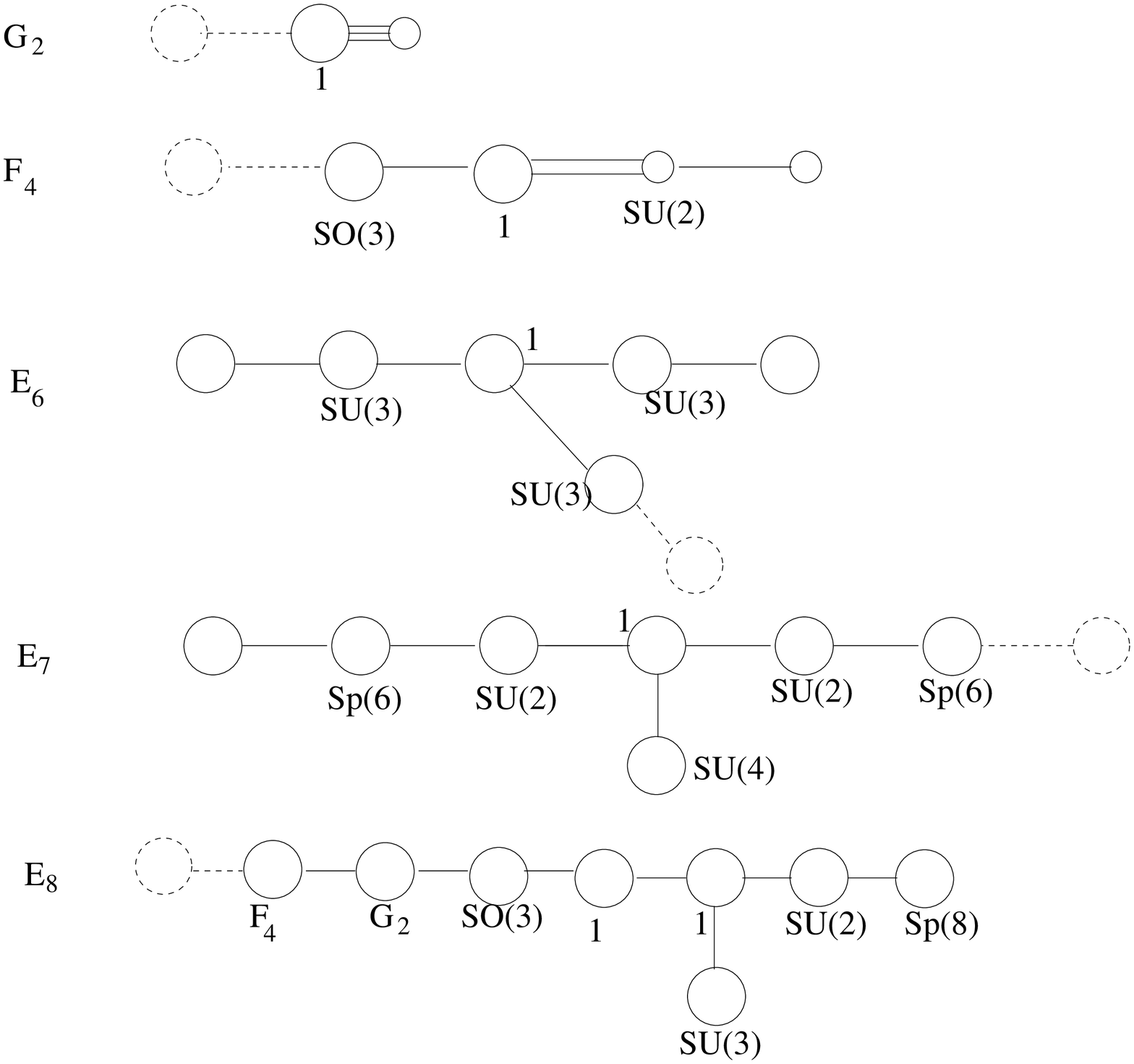}}
        \vspace{5mm}
    \end{center}
\caption{Centralizers of exceptional triples associated with a given node, 
$m = a_j^\lor$.}

\end{table}

\begin{table}
\label{tripcent78}

\begin{tabular}{||l|c|c|l||} 
$ G$ & node  &$ m$ & $G_{\rm triple}$ \\ \hline
$E_7$ & {\bf 4} & 2& $ [SU(2)]^3$ \\ \hline
$E_8$ & {\bf 6} & 2& $ [SU(3)]^2$ \\ \hline
$E_8$ & {\bf 4} & 2& $ SU(2) \times Spin(7)$ \\ \hline
$E_8$ & ${\bf 4}'$ & 2& $ SU(2) \times SU(4)$ \\ \hline
$E_8$ & {\bf 6} & 3& $ [SU(2)]^2$ \\ \hline
\end{tabular}

\caption{The same as in Table 7 with $m < a_j^\lor$. The node {\bf 4} for $E_8$
is placed on the ''long arm`` and the node ${\bf 4}'$ --- on the ''short arm``
 of the Dynkin diagram.}

\end{table}

In the case of $Spin(N)$ groups, these centralizers  are as follows :
 \be 
N &=& 2r + 1\ : \ G_{\rm triple}^{(k)} \ =\ Spin(2k - 1) \times
Spin(2r - 2k - 3),\ \  k = 1, \ldots, r-2  \nonumber \\  
N &=& 2r\ :\  \ G_{\rm triple}^{(k)} \ =\ Spin(2k - 1) \times
Spin(2r - 2k - 5),\ \  k = 1, \ldots, r-3  
 \ee
where $k$ labels the nodes with $a_i^\lor = 2$ on the Dynkin diagram of the
corresponding group. Note that in the case of even $N$ the maximal 
centralizer  is $G_{\rm triple}^{(1)} = G_{\rm triple}^{(r-3)} = Spin(N-7)$   
 as was noticed before in
 \cite{Witnew,KRS}. However, for $N = 2r + 1$, the maximal centralizer is
$Spin(N-6)$. An example of the corresponding non--trivial triple for $Spin(9)$
is $(\gamma_1 \gamma_2 \gamma_3 \gamma_4,\ \gamma_1 \gamma_2 \gamma_5, \ 
\gamma_1 \gamma_3 \gamma_6)$. It can be embedded in $Pin(6)$, but not in
$Spin(6)$. 

 For all
exceptional groups these centralizers are listed in   Table 7.
The groups $E_{7,8}$ involve also the nodes with not prime $a_j^\lor = a_j$.
These nodes admit triples with $m < a_j^\lor, \ m| a_j^\lor$. As was explained
in some details in Sect. 5, the centralizer of the 2--exceptional triple in
$E_7$ associated with the node {\bf 4} is $[SU(2)]^3$. The calculations for
$E_8$ are done in a similar way and the results are listed in Table 8.
Note that the centralizers of the triples with a given $m$ associated with 
different nodes might be different, but their rank is always the same as it
of course should be in virtue of Proposition 1.

\section*{Appendix B: Grassmann invariants of Weyl Groups. }

\setcounter{equation}0
\renewcommand{\theequation}{B.\arabic{equation}}

Let $ V = i\ Lie \ T$ be the euclidean $r$--dimensional vector space on which
the Weyl group $W$ acts via the reflection representation. We shall assume
in this section that the group $G$ is (almost) simple, hence the 
representation of $W$ in $V$ is irreducible. The group $W$ acts diagonally
on $V \oplus V$ and this action induces an action on the Grassmann algebra
$$ \Lambda \ =\ \Lambda (V \oplus V) \ . $$
Let $\Omega = \sum_i e_i^{(1)} \otimes  e_i^{(2)}$ be the obvious quadratic
invariant of $W$ in $V$, where $\{e_i^{(s)}\}^r_{i = 1}$, $s = 1$ or 2, is
the same orthonormal basis in both copies of $V$. Then $\Omega^j, \ \ 
j = 0, \ldots, r$, are linearly independent $W$--invariant elements of
$\Lambda$.

{\bf Theorem B1}. Elements $\Omega^j$, $j = 0, \ldots, r$, form a basis of
the space $\Lambda^W$ of $W$--invariants in $\Lambda$.

{\bf Proof}. It is a simple and well--known  fact that, for any 
representation of a finite group $\Gamma$ in a finite--dimensional vector space
$U$, the dimension of the space of invariants of $\Gamma$ in the Grassmann
algebra $\Lambda(U)$ over $U$ is given by the formula
  \be
\label{WAU}
{\rm dim}\ \Lambda (U)^\Gamma \ =\ \frac 1 {\# \Gamma} \sum_{g \in \Gamma}
\det (1 + g) \ .
 \ee
Applying this to $\Gamma = W$, $U = V \oplus V$, we see that Theorem B1 is
equivalent to the following formula:
  \be
\label{Weylinv}
\sum_{w \in W} [ P_w(-1)]^2 \ =\ (r + 1) \  \# W\ ,
 \ee
where $P_w(x)$ stands for the characteristic polynomial $\det (x - w)$ of $w$
acting on $V$. Fortunately, there is in Ref. \cite{C} a complete table of
 characteristic polynomials for all Weyl groups which allows one to check
(\ref{Weylinv}) in all cases. 

E. Vinberg pointed out that a conceptial
proof of  this theorem has been given by L. Solomon 
\cite{St} long ago:
 Note that dim $\Lambda^W \ =\ {\rm dim\ End}_W[\Lambda
(V)]$.  But all $\Lambda^j (V),\ j = 0,
 \ldots, 
r$, are irreducible and inequivalent representations of $W$. Hence the 
right--hand side of the last equality equals $r + 1$.  

{\bf Remark B1}. In fact, one has a refinement of (\ref{WAU}):
 \be
\label{WAUx}
\sum_{j \geq 0} {\rm dim}\  \Lambda^j (U)^\Gamma x^j \ =\ \frac 1 {\# \Gamma} 
\sum_{g \in \Gamma} \det (x + g) \ .
 \ee
Hence, Theorem B1 is equivalent to a refinement of (\ref{Weylinv}):
 \be
\label{Weylx}
1 + x^2 + x^4 + \ldots + x^{2r} \ =\ 
\frac 1 {\# W} 
\sum_{w \in W} [\det (x + g)]^2 \ .
  \ee
Therefore, letting $x = -1$, we have:
 \be
\label{B5}
 r + 1 \ = \ \frac 1 { \# W} \sum_{w \in W} \det (1 - w)^2 \ .
  \ee
The number $\det (1 - w)$ is called the defect of $w$. The defect of the
Coxeter elements is equal to $\#\ {\rm Center} (G)$. It follows from (\ref{B5})
that in the $A_n$ case ( and only in this case) all other elements have
defect 0 (cf. \cite{C}).

{\bf 
Corollary B1}.
The dimension of the space of the invariants of the group $W_{m}$ on the 
Grassmann algebra over $i \ Lie \ T_{m} \oplus i \ Lie \ T_{m}$ is 
$r_m + 1$.

{\bf Proof}.
In order to apply Theorem B1, we need to show that $W_{m}$ contains an 
irreducible finite reflection group. To show this, remark that if $T_{m}$
is a maximal torus of a compact subgroup, say $K$, of $G$, then  $W_{m}$
contains the Weyl group of $K$. From the list of groups  given in  Table
7, we conclude that  $W_{m}$ contains the subgroup listed
in the $W_{m}$ column of Table \ref{rmum}.

{\bf Remark B2.}
It is not difficult to show that the group listed in the $W_{m}$ column
of Table \ref{rmum} (denote this group by $W'_{m}$)
coincides with $W_{m}$. Indeed, we have:
\be
\label{WWprim}
W'_{m} \subseteq \ W_{m} \ \subseteq W \cap GL\left(i\ Lie\ T_{m}
\right) \ ,
 \ee
and one can check that in all cases the first group coincides with the last.
In the most difficult $E_8$ case, for $m=2$ (resp. 3), this follows from the 
fact that $W_{F_4}$ (resp. $W_{G_2}$ ) is a maximal finite subgroup of
$GL(m, {\bf Z})$. This argument works in all cases except for two:
{\it i)} $G = Spin(15) \ {\rm or} \ Spin(16)$ with 
$W_{m} = W_{B_4}$ which can in principle be embedded in $ W_{F_4}$ and
{\it ii)} $G = E_6, \ m=2$ with $W_{m} = W_{A_2}\ \subset W_{G_2}$. However,
the explicit calculations of Appendix C show that in all cases \ 
$\#W_{m}  = \#W'_{m}$. Together with Eq.(\ref{WWprim}), that means
that $W_{m}$ and $W'_{m}$ coinside.

{\bf Corollary B2.} If $W_m = W_{X_r}$ in Table 4, then the centralizer
of a lowest-dimensional $m$-exceptional triple in $G$ is a compact
Lie group of type $X_r$.

\section*{Appendix C: Tori, fundamental alcoves and triples' moduli space. }

\setcounter{equation}0
\renewcommand{\theequation}{C.\arabic{equation}}

We present here some explicit formulae for the fundamental weights $\omega_j$
entering the definition of the alcoves (\ref{conds}), (\ref{excel}), 
study the corresponding tori
(\ref{subsets}) and find the intersections $\Gamma_{m}$ of $T_{m}$ with the 
semi-simple parts of their centralizers $C_{m}$. 
We do it both by illustrative purposes and also 
because we need the groups $\Gamma_{m}$ to construct explicitly the moduli 
space (\ref{moduli}) of the commuting triples. 

We will not discuss here the groups $G_2$, $Spin(7)$, $Spin(8)$ where the
2--exceptional triples are isolated. That was done in enough details in the
main text.

\begin{figure}
\grpicture{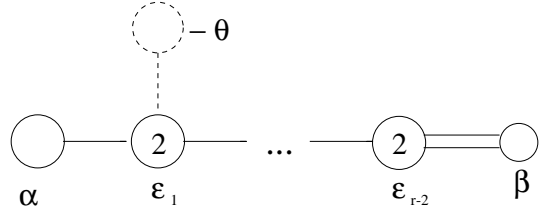}
\caption{Dynkin diagrams for $Spin(2r+1)$.}
\label{Dyn2r1}
\end{figure}

{\bf i)} $ Spin(2r+1),\ \ r > 3$.
The Dynkin diagram is drawn anew in Fig. \ref{Dyn2r1}. The fundamental 
alcove (\ref{conds}), (\ref{excel}) involves the weights $\omega_{\epsilon_1},
\ldots , \omega_{\epsilon_{r-2}}$ which can be conveniently written in this
case in the orthonormal 
basis $e_j = i \gamma_{2j-1} \gamma_{2j}/2$ :
 \be
\label{ves2r1}
\omega_{\epsilon_1}  \ &=&\ e_1 + e_2 \ =\ \theta^\lor \nonumber \\
\ldots \nonumber \\
\omega_{\epsilon_{r-2}}  \ &=&\ e_1 + e_2 + \ldots + e_{r-1}
  \ee
A general expression (\ref{excel}) for the exceptional element can be thus 
rewritten in the form $\sigma_{\vec{s}} = \exp\{\pi i(e_1 + e_2)\} t$, where
 \be
\label{t2r1}
t \ =\ \exp \left\{ \pi i (q_{r-3} e_3 + \ldots + q_1 e_{r-1}) \right \}
 \ee
with the condition $0 \leq q_1 \leq \ldots q_{r-3} \leq 1$. The volume of
the fundamental alcove is $V_{\rm dom} = 1/(r-3)!$. Taking $a_2 = \exp\{\pi i
\theta^\lor\}$ as the first element of the triple, we can pick up a Heisenberg
pair in $\tilde L_{2} = [SU(2)]^3$ formed by the roots $(\alpha, \beta, 
\theta)$. The canonical triple $(a_2, b_2, c_2)$ thus constructed lies in
$\tilde L_{2}$. It can be conjugated to the form
 \be
 \label{cantrip}
 a_2 \ =\   \gamma_1\gamma_2\gamma_3\gamma_4 \ ,  \ \ \ 
 b_2 \ =\   \gamma_1\gamma_2\gamma_{2r-1}\gamma_{2r} \ , \ \ \  
 c_2 \ =\ \gamma_1\gamma_3\gamma_{2r-1}\gamma_{2r+1}  \
  \ee
which coincides with Eq.(\ref{Spin7}) up to the change $5 \to 2r-1, 6
\to 2r, 7 \to 2r + 1$.
The fact that it can be embedded in $Spin(7)$ is seen quite 
directly, but an universal way to demonstrate this which works also for 
exceptional groups is to trade the root $\theta$ for
 \be
\label{zet2r1}
\zeta \ = \ \frac {\theta - \alpha - 2\beta}{2} \ = \ 
\epsilon_1 + \ldots + \epsilon_{r-2}
\ee
The roots $(\alpha, \zeta, \beta)$ present a system of simple roots for
$Spin(7)$.

The element (\ref{t2r1}) lies in the torus $T_{2}$ which commutes with
this $Spin(7)$. The torus presents a subset of the domain $\{0 \leq q_j
\leq 4 \}$ (indeed, $\exp\{4\pi i e_j\} = 1$). For $r > 4$, this domain
(one may call it {\it large} torus) should, however, be  factorized
by identifications $\{q_j\}:\ (0, \ldots, 0) \equiv (2,2, 0, \ldots, 0)
\equiv \ldots $ ( indeed, 
$ \exp\{2\pi i (e_3 + e_4)\} =  \ldots = 1$) .
As is not difficult to see, the corresponding factor group involves
$2^{r-4}$ elements and the volume of the torus is $V_{\rm tor}
= 4^{r-3}/2^{r-4} = 2^{r -2}$. Thereby, the fundamental alcove (\ref{t2r1})
is obtained from the torus by factorization with a finite group with
  \be
\frac {V_{\rm tor}}{V_{\rm dom}} \ =\ (r-3)!  \cdot 2^{r-2}
 \ee
elements. In the first place, this factorization is due to the action of the
Weyl group $W_{m}$ of the centralizer of our canonical triple. The 
continuous part of the latter is $Spin(2r - 6)$.
\footnote{This is so for the triple (\ref{cantrip}). As was mentioned before, 
one  can also choose a triple whose centralizer is larger $= \ Spin(2r-5)$. 
Such a triple cannot, however, be embedded in $C_2 = Spin(7)$ and is not 
convenient for our purposes.}
The corresponding Weyl group $W_{D_{r-3}}$ involves $(r-3)!  \cdot 2^{r-4}$
elements (the permutations of $e_j,\ \ j = 3,\ldots,r-1$ and the reflection
of a {\it pair} $e_{j_1} \to -e_{j_1},\ e_{j_2} \to -e_{j_2}$). However, a 
centralizer of our canonical triple  involves also an additional discrete
${\bf Z}_2$ factor. Indeed, the triple (\ref{cantrip}) commutes with the element
$\gamma_2\gamma_4\gamma_5\gamma_{2r}$ of the $Spin(2r+1)$ group. The 
conjugation of a generic triple by such an element amounts to the reflection
with respect to $e_3$: \ $q_{r-3} \to -q_{r-3}$.
  All together we obtain the group with $(r-3)!  \cdot 2^{r-3}$ elements 
involving all possible permutations and  reflections  of $e_j$, \ 
$j = 3,\ldots, r-1$. This is the group $ W_{B_{r-3}}$. Thus $W_m$
coincides with the Weyl group of the maximal centralizer of the triple
[$Spin(2r - 5)$ in this case] and we will see that this is also true in all 
other cases. 
 
 Besides, as
was explained in Sect. 4, there might be conjugations corresponding to a 
multiplication by a non--trivial element of the center of $Spin(7)$ which
belongs also to the torus. Indeed, the element $ \exp\{2\pi i e_3\} \equiv
\exp\{2\pi i e_4\} \equiv \ldots = -1$ of the torus belongs to the center of
$Spin(7)$ which is ${\bf Z}_2$ [in this simple case, it belongs also to the center
of the large group $Spin(2r + 1)$ , but this will not be so for the 
exceptional groups].

The elements of $W_{B_{r-3}}$ act on all elements of the triple simultaneously
while the elements of ${\bf Z}_2$ act on each component separately. We arrive 
thereby at the result (\ref{moduli}).

{\bf ii)} $ Spin(2r)$, $r > 4$.
All calculations are exactly parallel to the calculations for $Spin(2r + 1)$.
One only has to substitute $r \to r-1$ in all formulae. The canonical triple
is conveniently embedded in $Spin(8)$ ( and further in $Spin(7)$ ).
 There is an element
of the torus $T_{2}$ which coincides with the element $-1$ of the center
of $Spin(8)$. Another non--trivial element of Center($Spin(8)$), $\gamma_9
= \gamma_1  \cdots \gamma_8$, does not belong to the torus, however. Hence,
$\Gamma_2[Spin(2r)] = {\bf Z}_2$. In this case, the canonical triple has
a maximal centralizer $Spin(2r - 7)$ and $W_2 \ =\  W_{B_{r-4}}$.

\begin{figure}
\grpicture{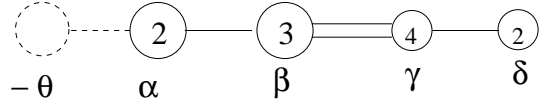}
\caption{Dynkin diagram for $F_4$.}
\label{DynF4}
\end{figure}

{\bf iii)} {$ F_4$}. 
The Dynkin diagram is depicted in Fig. \ref{DynF4}. 
Fisrt, there are two isolated 3--exceptional triples with the first element
$a_3 = \exp\left\{ \frac {2\pi i}3 \omega_\beta \right \}$, 
$\omega_\beta = 3\alpha + 6 \beta + 4 \gamma^\lor + 2 \delta ^\lor$.

Let us study the structure
of the fundamental alcove and of the torus $T_{2}$ associated with the
nodes with $a^\lor_j = 2$. We have
  \be
\label{vesF4}
\omega_{\alpha}  \ &=&\ 2\alpha^\lor + 3 \beta^\lor + 2 \gamma^\lor 
+ \delta^\lor \ =\ \theta^\lor \nonumber \\
\frac 12 \omega_{\gamma}  \ &=&\ 2\alpha^\lor + 4 \beta^\lor + 3 \gamma^\lor 
+ \frac 32 \delta^\lor 
  \ee 
A general $2$--exceptional element lying in the fundamental alcove
can be written in the form 
  \be
\label{domF4}
\sigma = \exp\{\pi i \theta^\lor\} \exp\{\pi i s \lambda\}\ ,
  \ee
 where
 $$
\lambda \ =\ \frac 12 \omega_\gamma - \omega_\alpha \ =\ 
\beta^\lor + \gamma^\lor + \frac 12 \delta^\lor 
 $$
and $0 \leq s \leq 1$.

The canonical triple with the first element $\exp\{\pi i \theta^\lor\}$
can be embedded in $[SU(2)]^3/{\bf Z}_2$ based on the roots $(\theta \beta \delta)$
and further in $Spin(7)$ with the simple roots $(\beta \zeta \delta)$, where
  \be
\label{zetF4}
\zeta \ = \ \frac {\theta - \beta - 2\delta}{2} \ = \ 
\alpha + \beta + 2\gamma
\ee 
Note that the highest root $\theta$ for $F_4$ is also the highest root for its
 $C_2$ subgroup $Spin(7)$. This property also holds for $Spin$ groups
[cf. Eq.(\ref{zet2r1})] and for higher exceptional groups.

The full torus $\exp\{\pi i  s \lambda\}$ corresponds to the domain $(0 \leq 
s \leq 4)$ ($\exp\{4 \pi i \lambda\} = 1$ due to the property 
$\exp\{2 \pi i \kappa^\lor \} = 1$ for any coroot $\kappa^\lor$) and is 4
times larger than the fundamental alcove (\ref{domF4}). The factorization
over ${\bf Z}_2$ comes from the Weyl group of
our 1--dimensional torus
  $W_{A_1} = {\bf Z}_2$ (with the non--trivial element acting as $s \to -s$).
Another ${\bf Z}_2 = \Gamma_{2}(F_4) $ corresponds to the shift $s \to s + 2$. Indeed,
$\exp\{ 2\pi i \lambda\} = \exp\{ \pi i \delta^\lor\} = -1$ as far as the 
$Spin(7)$  with the simple roots $(\beta \zeta \delta)$ is concerned.

\begin{figure}
\grpicture{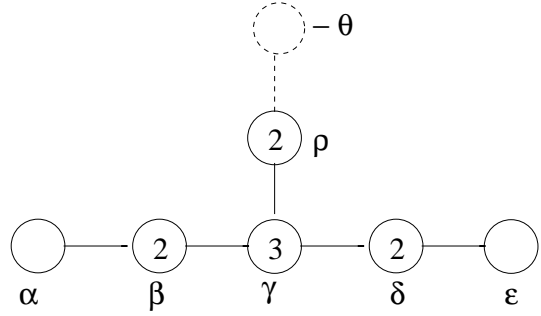}
\caption{Dynkin diagram for $E_6$.}
\label{DynE6}
\end{figure}

{\bf iv)} $E_6$.
 The Dynkin diagram is depicted in Fig. \ref{DynE6}. There are two isolated
3-exceptional triples associated with the node $\gamma$ with the first element 
   \be
 \label{a3F4}
a_3 \ = \ \exp\left\{ \frac {2\pi i}3 \omega_\gamma \right\}, \ \ \ 
\omega_\gamma \ =\ 2(\alpha^\lor + \epsilon^\lor ) + 
4(\beta^\lor + \delta^\lor) + 6 \gamma^\lor + 3 \rho^\lor 
  \ee

Note that  they can be embedded further in $F_4$. Indeed, the embedding
$F_4 \subset E_6$ is realized by gluing the nodes of the Dynkin diagram for
$E_6$ to its ''mirror images``: the root $\alpha$ is glued together with the 
root $\epsilon$ and the root $\beta$ together with the root $\delta$. To be 
quite exact, the coroots of the subgroup $F_4$ are expressed via the coroots
of $E_6$ as follows:
  \be
 \label{F4E6}
(\alpha^\lor, \beta^\lor, \gamma^\lor, \delta^\lor)_{\rm notation\ as\ in\ 
Fig.\ref{DynF4}} \  =\  (\rho^\lor, 
\gamma^\lor,  \beta^\lor  + \delta^\lor, \alpha^\lor + \epsilon^\lor)_{\rm 
notation\ as\ in\ Fig.\ref{DynE6}} 
 \ee
One can see directly that the 3--exceptional element (\ref{a3F4}) for $E_6$
is also the 3--exceptional element for $F_4$. Two other elements of the $E_6$
triple present the Heisenberg pairs in the $SU(3)$ factors formed by the nodes
$(\rho, -\theta)$, $(\alpha, \beta)$ and $( \delta, \epsilon)$. When going from
$E_6$ down to $F_4$, two latter subgroups are glued together so that the 
second and the third element of the $E_6$ triple coincide with the second and 
the third element of the $F_4$ triple.

Let us study now 2--exceptional triples associated with the nodes $\beta, 
\delta, \rho$ in $E_6$. We have
  \be
 \label{vesE6}
\omega_\beta \ &=&\ \frac 53 \left(  \alpha + 2 \beta \right)
+ 4\gamma + 2\rho + \frac 43 \left(  \epsilon + 2 \delta \right) \nonumber \\
\omega_\delta \ &=&\ \frac 43 \left(  \alpha + 2 \beta \right)
+ 4\gamma + 2\rho + \frac 53 \left(  \epsilon + 2 \delta \right) \nonumber \\
 \omega_\rho \ &=&\ \alpha + \epsilon + 2(\beta + \delta + \rho) + 3\gamma
= \theta
 \ee
 (as the groups $E_{6,7,8}$ are simply laced, we will not bother to mark 
coroots with the symbol ''$^\lor$`` anymore).

The fundamental alcove of 2--exceptional elements is 
   \be
  \label{domE6}
\sigma = \exp\{\pi i \theta\} \exp\{\pi i (s_1 \lambda_1 + s_2 \lambda_2)\} \ ,
\nonumber \\
s_1 + s_2 \ \leq 1, \ s_{1,2} \geq 0
\ ,
  \ee
where
 \be
 \label{lamE6}
\lambda_1 &=& \omega_\beta - \omega_\rho \ =\ \frac 13 (2\alpha + 4 \beta 
+ 3  \gamma + 2 \delta + \epsilon) \nonumber \\
\lambda_2 &=& \omega_\delta - \omega_\rho \ =\ \frac 13 (\alpha + 2 \beta 
+ 3  \gamma + 4 \delta + 2\epsilon) 
  \ee
We have 
$$ \langle  \lambda_1, \lambda_1 \rangle = \langle  \lambda_2, \lambda_2
 \rangle  = 4/3,\ \ \ \langle \lambda_1, \lambda_2 \rangle = 2/3 $$
so that the weights $\lambda_1, \lambda_2$ form the angle $\pi/3$.  

The canonical triple with the first element $\exp\{\pi i \theta^\lor\}$
can be embedded in $[SU(2)]^4/{\bf Z}_2$ based on the roots $(\theta, \alpha, \gamma,
 \epsilon)$
and further in $Spin(8)$ with the simple roots $(\alpha \gamma \epsilon)$ and
  \be
\label{zetE6}
\zeta \ = \ \frac {\theta - \alpha - \gamma - \epsilon}{2} \ = \ 
 \beta + \gamma + \delta + \rho 
\ee 
The root $\zeta$ lies in the center of the Dynkin diagram for $Spin(8)$
for which $\theta$ serves as the highest root.

\begin{figure}
\grpicture{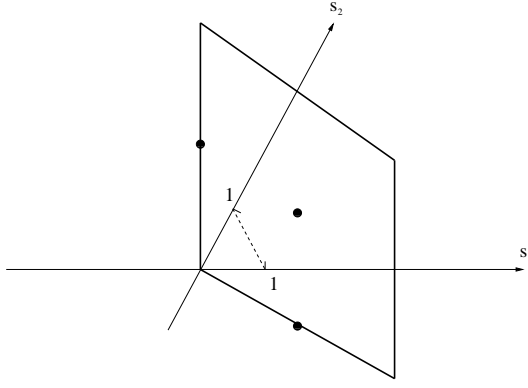}
\caption{Torus and fundamental alcove for $E_6$.}
\label{CellE6}
\end{figure}

The full torus corresponds to the rhombus  in Fig. \ref{CellE6}.
Indeed,
$$\exp\{2 \pi i (\lambda_1 + \lambda_2) \}  = 
\exp\{2 \pi i (2\lambda_1 - \lambda_2) \}  = 
\exp\{2 \pi i (2\lambda_2 - \lambda_1) \}  = \ldots = 1$$
( with the property 
$\exp\{2 \pi i \alpha_j \} = 1$ repeatedly used). It is 24 times larger than 
the fundamental alcove (the small triangle in Fig. \ref{CellE6} ). The 
factorization by 24 comes first from the Weyl group of the small torus 
$W_{A_2}$ ( \# $W_{A_2}$ = 6) and, second, from the group $\Gamma_2$ which
coincides in this case with the full center of $Spin(8)$ which is ${\bf Z}_2 \times
{\bf Z}_2$. Indeed, the elements 
 \be
 \label{elemE6}
\exp\{ \pi i (\lambda_1 + \lambda_2) \},\   
\exp\{ \pi i (2\lambda_1 - \lambda_2) \}, \ 
\exp\{ \pi i (2\lambda_2 - \lambda_1) \} 
  \ee
of the torus (marked out by the blobs in Fig. \ref{CellE6}) depend only on the
roots $\alpha, \ \gamma, \ \epsilon$ and belong therefore to our subgroup
$C_2(E_6) = Spin(8)$. As the whole torus commutes with this $Spin(8)$, the
elements (\ref{elemE6}) present 3 different non--trivial elements of its
center.

\begin{figure}
\grpicture{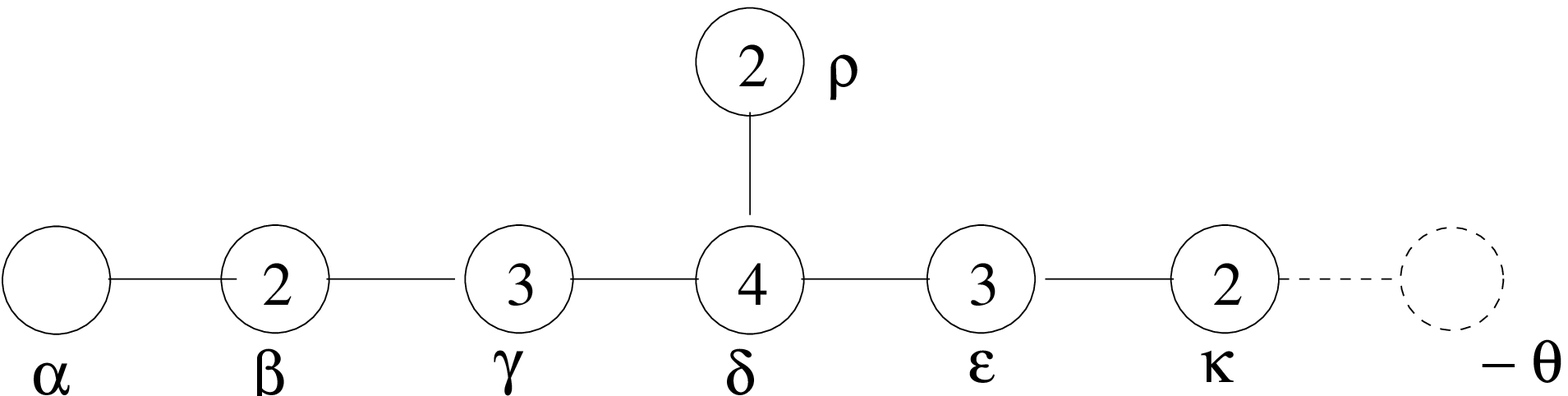}
\caption{Dynkin diagram for $E_7$.}
\label{DynE7}
\end{figure}

{\bf v)} $E_7$.
 The Dynkin diagram is depicted in Fig. \ref{DynE7}.
\begin{itemize}
\item $ m= 2$ . We have
  \be
 \label{vesE7}
\omega_\kappa \ &=&  \alpha + 2 \beta 
+ 3\gamma  + 4 \delta +  3\epsilon + 2\kappa + 2\rho  = \theta \nonumber \\
\omega_\beta \ &=& 2 \alpha + 4 \beta 
+ 5\gamma  + 6 \delta +  4\epsilon + 2\kappa + 3\rho  \nonumber \\
 \omega_\rho \ &=& \frac 32 \alpha + 3 \beta 
+ \frac 92 \gamma  + 6 \delta +  4\epsilon + 2\kappa + \frac 72 \rho  
\nonumber \\
\frac 12  \omega_\delta \ &=& \frac 32 \alpha + 3 \beta 
+ \frac 92 \gamma  + 6 \delta +  4\epsilon + 2\kappa + 3\rho  
     \ee
 The fundamental alcove  is 
\footnote{We hope that the universal notation $s_j$, $\lambda_j$ for 
{\it different} tori considered will bring
about no confusion.}
   \be
  \label{domE7}
\sigma = \exp\{\pi i \theta\} \exp\{\pi i (s_1 \lambda_1 + s_2 \lambda_2
+ s_3 \lambda_3)\} \ ,
\nonumber \\
s_1 + s_2 + s_3\ \leq 1, \ s_{1,2,3} \geq 0\ ,
  \ee
where
 \be
 \label{lamE7}
\lambda_1 &=& \omega_\beta - \omega_\kappa \ =\ \alpha + 2 \beta 
+ 2\gamma + 2\delta + \epsilon +  \rho 
\nonumber \\
\lambda_2 &=& \omega_\rho - \omega_\kappa \ =\ \frac 12 \alpha +  \beta 
+ \frac 32  \gamma + 2 \delta + \epsilon + \frac 32 \rho  \nonumber \\
\lambda_3 &=& \frac 12 \omega_\delta - \omega_\kappa \ =\ \frac 12 \alpha +  
\beta + \frac 32 \gamma + 2 \delta + \epsilon + \rho
  \ee

The canonical triple with the first element $\exp\{\pi i \theta\}$
can be embedded in $[SU(2)]^4/{\bf Z}_2$ 
and further in $Spin(8)$ with the simple roots $(\alpha \gamma \epsilon)$ and
  \be
\label{zetE7}
\zeta \ = \ \frac {\theta - \alpha - \gamma - \epsilon}{2} \ = \ 
 \beta +  \gamma  + 2\delta +  \epsilon +   \kappa + \rho 
\ee 
 The  torus corresponds to the domain $(0 \leq s_1 \leq 2, \ \ 0 \leq s_{2,3}
\leq 4)$ (without further factorizations).  The ratio
\be
 \label{ratE7}
\frac{V_{\rm tor}}{V_{\rm alc}} \ =\ \frac {32}{1/6} \ =\ 4 \# W_{C_3}
 \ee
is brought about by the factorization over the Weyl group of $Sp(6)$ , the 
maximal centralizer of the triple ( see Table 7; the Weyl group of $Sp(6)$
is the same as for the dual $Spin(7)$ and consists of $2^3 \cdot 3! 
= 48$ elements), and the factorization over $\Gamma_2(E_7) = {\bf Z}_2 \times {\bf Z}_2$.
 Three non--trivial elements of $\Gamma_2(E_7)$ are the following elements  of
 the torus:
 \be
 \label{elemE7}
\exp\{2 \pi i \lambda_3  \} &=& \exp\{ \pi i (\alpha + \gamma) \} \nonumber \\
\exp\{ \pi i (\lambda_1 + 2 \lambda_2 ) \} & =& \exp\{ \pi i ( \gamma +
\epsilon)  \} \nonumber \\ 
\exp\{ \pi i (\lambda_1 + 2\lambda_2 + 2\lambda_3) \} & =& \exp\{ \pi i
 ( \alpha + \epsilon)  \}
  \ee

\item $m = 3$. The relevant nodes are $\gamma$ and $\epsilon$. 
 We have
  \be
 \label{vesE7m3}
\omega_\gamma \ &=&  \frac 52 \alpha + 5 \beta 
+ \frac {15}2 \gamma  + 9 \delta +  6\epsilon + 3\kappa + \frac 92 \rho  
 \nonumber \\
\omega_\epsilon \ &=& 2 \alpha + 4 \beta 
+ 6\gamma  + 8 \delta +  6\epsilon + 3\kappa + 4\rho  
     \ee
 The fundamental alcove  is 
   \be
  \label{domE7m3}
\sigma = \exp\{(2\pi i/3) \omega_\epsilon \} \exp\{(2\pi i/3) s \lambda \} \ ,
\ \ \ 0 \leq s \leq 1 \ ,
  \ee
where
 \be
 \label{lamE7m3}
\lambda = \omega_\gamma - \omega_\epsilon \ =\ \frac 12 \alpha +  \beta 
+ \frac 32 \gamma + \delta + \frac 12  \rho 
  \ee
The canonical triple with the first element $\exp\{(2\pi i/3) 
\omega_\epsilon\}$ can be embedded in $[SU(3)]^3/{\bf Z}_3$ 
and further in $E_6$ with the simple roots $(\alpha \beta, \rho\delta, 
\kappa )$ and the root
  \be
\label{zetE7m3}
\zeta \ = \ \frac {\theta - \alpha - \delta - 2(\beta + \rho + \kappa)}{3}
 \ = \ \gamma + \delta  + \epsilon 
\ee 
with nonzero projections $ \langle \zeta, \beta \rangle =  \langle \zeta, 
\rho \rangle =  \langle \zeta, \kappa \rangle
= -1$.

 The  torus corresponds to the domain $(0 \leq s \leq 6)$. It gives the
fundamental alcove after factorization over $W_{A_1} = {\bf Z}_2$ and over
the group $\Gamma_3(E_7) = {\bf Z}_3$.
 One can see, indeed,  that the  elements 
$\exp\{(4\pi i/3)  \lambda \}$ and $\exp\{(8\pi i/3)  \lambda \}$ 
 of the torus in Eq.(\ref{domE7m3}) belong to $E_6$ constructed above and to
its center.

\item There are also two isolated triples with $m = 4$.
\end{itemize}

\begin{figure}
\grpicture{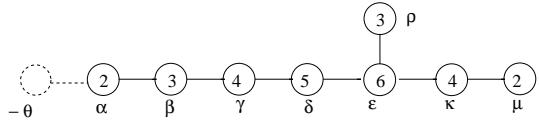}
\caption{Dynkin diagram for $E_8$.}
\label{DynE8}
\end{figure}

{\bf vi)} $E_8$.
 The Dynkin diagram is depicted in Fig. \ref{DynE8}.
\begin{itemize}
\item $ m = 2$ .  This is the most rich and beautiful case. 
There are 5 relevant nodes: $\alpha, \gamma, \epsilon, \kappa, \mu$. We have
  \be
 \label{vesE8}
\omega_\alpha \ &=& 2 \alpha + 3 \beta 
+ 4\gamma  + 5 \delta +  6\epsilon + 4\kappa + 2\mu + 3\rho  = \theta 
\nonumber \\
\frac 12 \omega_\gamma \ &=&  2\alpha + 4 \beta 
+ 6\gamma  + \frac {15}2 \delta +  9 \epsilon + 6 \kappa +  3 \mu + 
\frac 92\rho \nonumber \\
 \frac 12 \omega_\kappa \ &=&  2\alpha + 4 \beta 
+ 6\gamma  + 8 \delta +  10 \epsilon + 7 \kappa +  \frac 72 \mu + 5\rho
 \nonumber \\
 \frac 13 \omega_\epsilon \ &=&  2\alpha + 4 \beta 
+ 6\gamma  + 8 \delta +  10 \epsilon + \frac {20}3  \kappa + \frac {10}3 \mu +
  5\rho  \nonumber \\
   \omega_\mu \ &=&  2\alpha + 4 \beta 
+ 6\gamma  + 8 \delta +  10 \epsilon + 7 \kappa +  4 \mu + 5\rho
 \nonumber \\
     \ee
The fundamental alcove  is 
   \be
  \label{domE8m2}
\sigma = \exp\{\pi i \theta\} \exp\{\pi i (s_1 \lambda_1 + s_2 \lambda_2
+ s_3 \lambda_3 + s_4 \lambda_4)\} \ ,
\nonumber \\
s_1 + s_2 + s_3 + s_4\ \leq 1, \ s_{1,2,3,4} \geq 0\ ,
  \ee
where
 \be
 \label{lamE8m2}
\lambda_1 &=& \frac 12 \omega_\gamma - \omega_\alpha \ =\  \beta 
+ 2\gamma + \frac 52 \delta + 3\epsilon +  2\kappa + \mu + \frac 32 \rho 
\nonumber \\
\lambda_2 &=& \frac 13 \omega_\epsilon - \omega_\alpha \ =\ 
\beta 
+ 2\gamma + 3 \delta + 4\epsilon +  \frac 83 \kappa + \frac 43 \mu + 2 \rho 
  \nonumber \\
\lambda_3 &=& \frac 12 \omega_\kappa - \omega_\alpha \ =\ 
\beta + 2\gamma + 3 \delta + 4\epsilon +  3 \kappa + \frac 32 \mu + 2 \rho 
\nonumber \\
\lambda_4 &=&  \omega_\mu - \omega_\alpha \ =\ 
\beta + 2\gamma + 3 \delta + 4\epsilon +  3 \kappa + 2 \mu + 2 \rho
\ee
The canonical triple with the first element $\exp\{\pi i \theta\}$
can be embedded in $[SU(2)]^4/{\bf Z}_2$ 
and further in $Spin(8)$ with the simple roots $(\beta \delta \rho)$ and the
central root
  \be
\label{zetE8m2}
\zeta \ = \ \frac {\theta - \beta - \delta - \rho}{2} \ = \ 
 \alpha + \beta +  2\gamma  + 2\delta +  3\epsilon +  2 \kappa + \mu + \rho 
\ee 
 The  torus corresponds to the domain $(0 \leq s_{1,3} \leq 4, \ 
\ 0 \leq s_{2}
\leq 6,\ \ 0 \leq s_4 \leq 2)$.  The ratio
\be
 \label{ratE8}
\frac{V_{\rm tor}}{V_{\rm alc}} \ =\ \frac {192}{1/24} \ =\ 4 \# W_{F_4}
 \ee
is brought about by the factorization over the Weyl group of the
maximal centralizer of the triple $F_4$ 
($\# W_{F_4} = 1152$)
and over $\Gamma_2(E_8) = {\bf Z}_2 \times {\bf Z}_2$.
 Three non--trivial elements of $\Gamma_2(E_8)$ are the following elements  of
 the torus:
 \be
 \label{elemE8}
\exp\{2 \pi i \lambda_1  \} &=& \exp\{ \pi i (\delta  + \rho) \} \nonumber \\
\exp\{3 \pi i   \lambda_2  \} & =& \exp\{ \pi i ( \beta +
\delta)  \} \nonumber \\ 
\exp\{ \pi i (2\lambda_1 + 3\lambda_2) \} & =& \exp\{ \pi i
 ( \beta + \rho)  \}
  \ee

\item $m = 3$. The relevant nodes are $\beta$,  $\epsilon$ and $\rho$. 
The corresponding fundamental weights are
  \be
 \label{vesE8m3}
\omega_\beta \ &=&  3 \alpha + 6 \beta 
+ 8 \gamma  + 10 \delta +  12\epsilon + 8\kappa +  4\mu + 6 \rho  
 \nonumber \\
\frac 12 \omega_\epsilon \ &=& 3 \alpha + 6 \beta 
+ 9 \gamma  + 12 \delta +  15\epsilon + 10\kappa +  5\mu + \frac{15}2  \rho 
  \nonumber \\
\omega_\rho \ &=&  3 \alpha + 6 \beta 
+ 9 \gamma  + 12 \delta +  15\epsilon + 10\kappa +  5\mu + 8 \rho  
     \ee
 The fundamental alcove  is 
   \be
  \label{domE8m3}
\sigma = \exp\{(2\pi i/3) \omega_\beta \} \exp\{(2\pi i/3) (s_1 \lambda_1
+ s_2 \lambda_2 \} \ , \nonumber \\
s_{1,2} \geq  0,\ \ s_1 + s_2 \leq 1 \ ,
  \ee
where
 \be
 \label{lamE8m3}
\lambda_1\  &=& \ \frac 12 \omega_\epsilon - \omega_\beta \ =\ 
  \gamma + 2\delta + 3\epsilon + 2\kappa + \mu + \frac 32  \rho 
\nonumber \\
\lambda_2 \ &=& \  \omega_\rho - \omega_\beta \ =\ 
 \gamma + 2\delta + 3\epsilon + 2\kappa + \mu + 2 \rho 
  \ee
The canonical triple with the first element $\exp\{(2\pi i/3) \omega_\beta\}$
can be embedded in $[SU(3)]^3/{\bf Z}_3$ 
and further in $E_6$ with the simple roots $(\alpha, \gamma\delta, 
\mu \kappa )$ and the root
  \be
\label{zetE8m3}
\zeta \ = \ \frac {\theta - \gamma - \kappa - 2(\alpha +  \delta + \mu )}{3}
 \ = \ \beta + \gamma + \delta  + 2\epsilon + \kappa  + \rho 
\ee 
with nonzero projections $\langle \zeta, \alpha\rangle  = \langle \zeta, 
\delta\rangle  = \langle \zeta, \mu\rangle 
= -1$.

 The  torus corresponds to the domain $(0 \leq s_1 \leq 6, \ \ 0 \leq
s_2 \leq 3)$. Its volume is 36 times larger than the volume of the fundamental
alcove (\ref{domE8m3}). The latter is obtained from the torus    after 
factorization over $W_{G_2}$ (the Weyl group for $G_2$ has 12 elements
involving besides  the elements from $W_{A_2}$ also 6 elements
 $-W_{A_2}$) and further over the group $\Gamma_3(E_8) = {\bf Z}_3$.
Indeed,   the  elements 
$\exp\{(4\pi i/3)  \lambda_1 \}$ and $\exp\{(8\pi i/3)  \lambda_1 \}$ 
 of the torus  belong to $E_6$.

 \item $m = 4$. The relevant nodes are $\gamma$ and $\kappa$. 
We have
  \be
 \label{vesE8m4}
\omega_\gamma \ &=&  4 \alpha + 8 \beta 
+ 12 \gamma  + 15 \delta +  18\epsilon + 12\kappa +  6\mu + 9 \rho  
 \nonumber \\
 \omega_\kappa \ &=& 4 \alpha + 8 \beta 
+ 12 \gamma  + 16 \delta +  20\epsilon + 14\kappa +  7\mu + 10  \rho 
     \ee
 The fundamental alcove  is 
   \be
  \label{domE8m4}
\sigma = \exp\{(\pi i/2) \omega_\gamma \} \exp\{(\pi i/2) s \lambda
 \} \ ,\ \ \ \ \ \  0 \leq s \leq 1 \ ,
  \ee
where
 \be
 \label{lamE8m4}
\lambda\  = \  \omega_\kappa - \omega_\gamma \ =\   \delta + 2\epsilon + 
2\kappa + \mu +  \rho 
  \ee
The canonical triple with the first element $\exp\{(\pi i/2) \omega_\gamma\}$
can be embedded in $[SU(4) \times SU(4) \times SU(2)]/{\bf Z}_4$ 
and further in $E_7$ with the simple roots $(\alpha\beta, \mu, \rho
\epsilon\delta )$ and the root
  \be
\label{zetE8m4}
\zeta \ = \ \frac {\theta - \delta - 2(\alpha +  \epsilon + \mu ) 
- 3(\beta + \rho)}{4}
 \ = \ \gamma + \delta  + \epsilon + \kappa 
\ee 
with nonzero projections $\langle \zeta, \beta\rangle  = \langle \zeta, 
\rho\rangle  = \langle \zeta, \mu\rangle 
= -1$.

 The  torus corresponds to the domain $(0 \leq s \leq 4)$. 
It is four times larger than the alcove in Eq.(\ref{domE8m4}). 
 The latter is obtained from the torus    after 
factorization over $W_{A_1} = {\bf Z}_2$  and further over the group 
$\Gamma_4(E_8) = {\bf Z}_2$.
Indeed,  $\exp\{\pi i \lambda\} = \exp\{\pi i (\delta + \mu + \rho)\}
\in E_7$.

\item And there are, of course, also ''genuine`` $E_8$ triples with $m = 5$
and $m = 6$ which are isolated.

\end{itemize}

\section*{Appendix D:  Minimal non--trivial $n$--tuples. }

\setcounter{equation}0
\renewcommand{\theequation}{D.\arabic{equation}}

The main subject of this paper were the non--trivial commuting triples. But one
can equally well be interested in non--trivial quadruples, quintuples etc.
If we define a non--trivial $n$--tuple as an (ordered)  set of $n$ commuting 
elements which cannot be simultaneously conjugated to the maximal torus, 
they are abundant in any
group which admits a nontrivial triple: just pick up such a triple and complete
it with arbitrary $n-3$ commuting elements from the centralizer of this triple.

It is clear that the classification of non--trivial commuting $n$--tuples 
is reduced to the classification of  $n$--tuples of a special kind which
we will call {\it minimal}:

{\bf Definition}. A minimal commuting $n$--tuple is a  non--trivial commuting
$n$--tuple such that its any $m$--element subset with $m < n$
 is a trivial $m$--tuple.

{\bf Example D1} \cite{Vinb} . Consider the following set of 4 elements of
$SO(16)$:

\be
\label{Vinb4}
\Omega_1 \ = \ {\rm diag} (-1,-1,-1,-1,-1,-1,-1,-1,1,1,1,1,1,1,1,1) 
\nonumber \\
\Omega_2 \ = \ {\rm diag} (-1,-1,-1,-1,1,1,1,1,-1,-1,-1,-1,1,1,1,1) 
\nonumber \\ 
\Omega_3 \ = \ {\rm diag} (-1,-1,1,1,-1,-1,1,1,-1,-1,1,1,-1,-1,1,1) 
\nonumber \\
\Omega_4 \ = \ {\rm diag} (-1,1,-1,1,-1,1,-1,1,-1,1,-1,1,-1,1,-1,1) 
\ee
 This quadruple is a natural generalization of the non--trivial triple in 
$Spin(8)$ written in \cite{Witnew}: 
write a $4 \times 16$ matrix whose 
 columns are all possible 4--vectors with coordinates 
$\pm 1$ taken 
once; then the 4 rows of this matrix viewed as diagonal matrices in $SO(16)$
is the quadruple (\ref{Vinb4}).

Obviously, $[\Omega_i, \Omega_j] = 0$, and also the liftings of $\Omega_i$ 
 up to $Spin(16)$:
 \be
\label{4Spin16}
 \tilde \Omega_1 \ &=&\ \gamma^1 \gamma^2 \gamma^3 \gamma^4 \gamma^5 \gamma^6 
\gamma^7 \gamma^8 \nonumber \\
 \tilde \Omega_2 \ &=&\ \gamma^1 \gamma^2 \gamma^3 \gamma^4 \gamma^9 
\gamma^{10} \gamma^{11} \gamma^{12} \nonumber \\
 \tilde \Omega_3 \ &=&\ \gamma^1 \gamma^2 \gamma^5 \gamma^6  \gamma^9 
\gamma^{10} \gamma^{13} \gamma^{14} \nonumber \\
   \tilde \Omega_4 \ &=&\ \gamma^1 \gamma^3 \gamma^5 \gamma^7 \gamma^9 
\gamma^{11} \gamma^{13} \gamma^{15}
 \ee
[a direct generalization of Eq.(\ref{Spin7})] commute. 

The quadruple (\ref{4Spin16}) is non--trivial. Indeed, consider the centralizer
$G_{\tilde \Omega_1} \ =\ [Spin(8) \times Spin(8)]/{\bf Z}_2$. In this case, the 
presence of the ${\bf Z}_2$ factor over which the product $Spin(8) \times Spin(8)$
is factorized is irrelevant. What is relevant is that each $Spin(8)$ factor
admits a non--trivial triple. It is not difficult to see that 
$( \tilde \Omega_2, \  \tilde \Omega_3, \  \tilde \Omega_4)$ present indeed
a non--trivial triple in {\it each} $Spin(8)$ factor. Therefore, 
$( \tilde \Omega_2, \  \tilde \Omega_3, \  \tilde \Omega_4)$ cannot be 
conjugated to the maximal torus in $G_{\tilde \Omega_1}$ and the whole 
quadruple (\ref{4Spin16}) cannot be conjugated to the maximal torus in
$Spin(16)$. 

On the other hand, any pair like $( \tilde \Omega_2, \  \tilde \Omega_3)$ can 
be
conjugated to the maximal torus in $G_{\tilde \Omega_1}$ and hence the triple
$( \tilde \Omega_1, \  \tilde \Omega_2, \  \tilde \Omega_3)$ can be
conjugated to the maximal torus in $G$. Also the triple 
$( \tilde \Omega_2, \  \tilde \Omega_3, \  \tilde \Omega_4)$ is trivial. One 
can be directly convinced that it can be conjugated to the
maximal torus in $Spin(16)$ involving the generators 
 \be
\label{tormix}
T_{1,9}, \ldots , T_{8,16}\ .
  \ee
 This torus kind of ``mixes'' two $Spin(8)$ 
factors in $G_{\tilde \Omega_1}$. 

Thereby, the quadruple (\ref{4Spin16}) is a minimal quadruple.
Since $Spin(16)$ involves a unique up to conjugacy
 element whose centralizer involves {\it
two} $Spin(8)$ factors and since there is a unique up to conjugacy
 triple 
$( \tilde \Omega_2, \  \tilde \Omega_3, \  \tilde \Omega_4)$ 
in $Spin(8) \times Spin(8)$ which is non--trivial in both factors, it follows 
that all minimal quadruples in $Spin(16)$ are equivalent by conjugation to
Eq.(\ref{4Spin16}):
the quadruple (\ref{4Spin16}) is isolated. 
The quadruple (\ref{4Spin16}) does not involve
$\gamma^{16}$ and presents obviosly also a minimal isolated
quadruple in  $Spin(15)$ [its image in  $SO(15)$ coincides with 
Eq.(\ref{Vinb4}) with the last column crossed out]. 

{\bf Example D2}. The construction of Example D1 is easily generalized to
 minimal $n$--tuples with arbitrary $n \geq 4$.
Write a $4 \times 2^n$ matrix whose 
 columns are all possible $n$--vectors with coordinates 
$\pm 1$ taken 
once and treat the rows as a set of $n$   diagonal $SO(2^n)$ matrices.  
 It is not 
difficult to see
that the set of the corresponding $n$ elements in   $Spin(2^n)$ presents a 
minimal $n$--tuple.
Indeed, take one of such elements  $\tilde \Omega_1$ in $Spin(2^n)$
 presenting a product of some $2^{n-1}$
gamma--matrices. The centralizer $G_{\tilde \Omega_1}$ is   
$ [Spin(2^{n-1}) \times Spin(2^{n-1})]/{\bf Z}_2$. One can see by induction that 
the
set $\tilde \Omega_2, \ldots, \tilde \Omega_n$ of remaining $n-1$ elements of our set
 presents the minimal $(n-1)$--tuple in each
$Spin(2^{n-1})$ factor. Therefore this set is not conjugable to the maximal 
torus in the centralizer and the
whole set of $n$ elements in not conjugable to the maximal torus in 
$Spin(2^n)$. This non--trivial $n$--tuple is minimal:
any set of $n-1$ elements including $\tilde \Omega_1$ is conjugable to the maximal 
torus due to the fact that the
$(n-1)$--tuple  $\tilde \Omega_2, \ldots, \tilde \Omega_n$ is minimal   in each
simple  factor of  $\tilde  G_{\tilde \Omega_1}$, and also the set $\tilde \Omega_2, \ldots, 
\tilde \Omega_n$ is conjugable to the
maximal torus in $Spin(2^n)$: this torus is formed by the generators 
$T_{1,2^{n-1} + 1}, \ldots, T_{2^{n-1}, 2^n}$. 

The constructed $n$--tuple in
$Spin(2^n)$
is isolated and presents  obviously also an isolated $n$--tuple in
$Spin(2^n-1)$.

 The minimal $n$--tuples exist also for $Spin(N)$ groups with
$N = 2^n + 2s$ and $N = 2^n + 2s-1$, $s > 0$, but they are not isolated 
anymore, a 
moduli space appears. Consider
a subgroup $Spin(2^n) \subset Spin(N)$ and pick up the minimal $n$--tuple 
$\tilde \Omega_1, \ldots, \tilde \Omega_n$
in $Spin(2^n)$. Its centralizer is a group of rank $s$. Let its maximal torus
be $T^s$. The  set of elements
$\tilde \Omega_1 t_1, \ldots, \tilde \Omega_n t_n$ , $t_j \in T^s$ presents 
also a  
minimal $n$--tuple. Thus, for $N > 2^n$ the moduli space of
minimal $n$--tuples is a subset of  
$ \tilde \Omega_1 T^s \times \cdots  \times \tilde \Omega_n T^s$. Moreover, 
one can easily generalize   the 
analysis of Section 4 and Appendix
C  to this 
case and find the moduli space exactly. We have  
 \be
\label{mod4Spin}
{\cal M}_{n-{\rm tuple}}^{Spin(2^{n} + 2s - 1)} \ = \ 
{\cal M}_{n-{\rm tuple}}^{Spin(2^{n} + 2s)} \ = \ \left.
\left[ \frac{  \tilde \Omega_1 T^{s}}{{\bf Z}_2} \times 
\cdots \times  \frac{  \tilde \Omega_n T^{s}}{{\bf Z}_2} 
\right] \right/ W_{B_{s}} \ ,
 \ee
where $W_{B_{s}}$ is the Weyl group of the maximal centralizer of a minimal
$n$--tuple in $Spin(N)$ [this centralizer  is $Spin(2s+1)$] and ${\bf Z}_2$ is the 
intersection of $T^s$
with the center of $Spin(2^n)$ or $Spin(2^n-1)$ for even and odd $N$, 
respectively.
\footnote{The groups $Spin(2^n)$ or $Spin(2^n-1)$ are the centralizers of
$T^s$ in large $Spin(N)$ group and play the role of the subgroups $C_m$
listed in Table 4 which were relevant when discussing the moduli space of
non--trivial commuting triples. }

{\bf Example D3}. Consider the group $E_8$ and take the element  
$g = \exp\{i\pi \omega_\mu\}$ 
[the notations are as in Fig. \ref{DynE8}]. By Theorem 1, 
its centralizer 
is $Spin(16)/{\bf Z}_2$. Take the minimal quadruple (\ref{4Spin16}) in $Spin(16)$. 
Denote by $Q$  the corresponding set
of  $E_8$ elements. 
The same reasoning as before shows that the quintuple $(g, Q)$ is non--trivial.
 On the other hand,
we will see later that $E_8$ does not admit minimal quadruples which means that
 $Q$ is trivial in $E_8$. It follows then that our quintuple is minimal.
Note that, as is also the case for the minimal quadruple (\ref{4Spin16}) and
all non--trivial triples and minimal $n$--tuples considered before, the first 
element $g$ in the
quintuple $(g, Q)$ does not play a special role: all other elements of this 
quintuple are equivalent to $g$ by conjugation.

{\bf Theorem D1}. 
{\it (a)}
 $SU(N)$ and $Sp(N)$ contain no 
non--trivial $n$--tuples.

{\it (b)}
$Spin(N)$ for $N < 15$ and the groups $G_2,\ F_4,\ E_6,\ E_7$ 
 contain no  minimal $n$--tuples for $n > 3$.

{\it (c)} 
$E_8$ contains no minimal $n$--tuples for $n=4$ and $n > 5$. $E_8$ contains 
a unique up to conjugacy  minimal quintuple and it is described in Example D3.

{\it (d)}
 All minimal $n$--tuples with $n > 3$ in  $Spin(N)$, $N \geq 15$
are described in Example D2.

{\bf Proof.}
{\it (a)} is well known (and follows from Theorem 1c). Next, 
 if $a_1$ is an element of a minimal $n$--tuple
$(a_1, \ldots, a_n)$ in $G$ with $n >3$, then at least one of the simple
 components of
$ G_{a_1}$ should not be of $A$ or $C$ type. Indeed, otherwise either
the preimage $(\tilde a_2, \ldots, \tilde a_n)$ in $\tilde G_{a_1}$ is
a trivial $(n-1)$--tuple in which case the whole $n$--tuple is trivial in $G$
[due to {\it (a)}] 
or some pair $\tilde a_i$, $\tilde a_j$ ($1 < j < j \leq n$) does not commute 
in which
case the triple $(a_1, a_i, a_j)$ is non--trivial contradicting the minimality
assumption. 

{\it (b)}. The centralizer of any element of $Spin(7)$ or $Spin(8)$ not 
belonging to the center presents a product of simple factors of $A$ or $C$ type
and the remark above applies. Consider $Spin(9)$ and suppose that it 
contains a a minimal $n$--tuple
$(a_1, \ldots, a_n)$ with $n \geq 4$. By the same remark, the only 
possibilities
for the semi--simple part of $G_{a_1}$ ( and for all  $G_{a_i}$ as well) to
be discussed are
$Spin(7)$ or $Spin(8)$ (see Theorem 1) and $(a_2, \ldots, a_n)$ should be a
non--trivial $(n-1)$--tuple in $G_{a_1}$. We have just shown that $Spin(7)$ and
 $Spin(8)$ do not admit minimal $n$--tuples with $n \geq 4$, hence $G_{a_1}$ 
should contain
a non--trivial triple. A non--trivial triple of $Spin(8)$ presents also a
non--trivial triple in its subgroup $Spin(7)$.  But it follows from Appendix C 
and Section 
4 that the isolated triple $(a,b,c)$ in $Spin(7)$ remains  non--trivial in 
$Spin(9)$ and that all non--trivial triples in $Spin(9)$ are up to conjugacy
of the form
 \be
\label{cantrip1}
(a t_1, b t_2, c t_3)\ ,
 \ee
where $t_i$ lie in a maximal torus of the centralizer of $Spin(7)$ in
 $Spin(9)$. Hence our $n$--tuple is not minimal.

Consider $Spin(N)$, $9 < N < 15$. The centralizer of any element of such
a group can involve only {\it one} simple factor which is not of $A$
or $C$ type, and this factor is  of $Spin(M < N)$ (i.e.  $B$ or $D$) type.
For an $n$--tuple  $(a_1, \ldots, a_n)$ to be non--trivial, the 
$(n-1)$--tuple  $(a_2, \ldots, a_n)$ should be non--trivial in  $G_{a_1}$.
Using the previous results and  inductive reasoning, we deduce that  $G_{a_1}$
 and its $Spin(M)$ factor should contain a non--trivial triple. Again, the
analysis of Appendix C and Section 4 shows that any such  triple is of
the form (\ref{cantrip1}) where $(a,b,c)$ is a  canonical triple built up 
according to the rules of
Appendix C which can be embedded in $Spin(7)$, and  $t_{i}$ lie
 on the maximal torus of the centralizer of the
canonical triple [and $Spin(7)$] in $Spin(M)$. The same applies to $Spin(N)$.
These maximal tori can be embedded one into another and hence a non--trivial
triple of $Spin(M)$ is also a non--trivial triple in $Spin(N)$.  

Let now $G$ be one of the exceptional groups. A centralizer of any element
in $G_2$ involves only semi--simple factors of $A$ or $C$ type and hence
$G_2$ does not contain a non--trivial $n$--tuple with $n > 3$. Consider 
$F_4$. There are elements with centralizer involving the factor $Spin(7)$
or $Spin(9)$. Again, the construction of the Appendix C and the
reasoning above can be used to show
that any non--trivial triple of $Spin(7)$ or $Spin(9)$ remains also 
non--trivial in $F_4$. 
In the $E_6$ case, the only possible 
semi--simple factors in the centralizer which are not of  $A$ or $C$ type 
are $Spin(8)$ or $Spin(10)$, and we argue in the same way as for $Spin(9)$
and $F_4$. 
In the $E_7$ case, the only possible 
semi--simple factors in the centralizer which are not of  $A$ or $C$ type 
are $Spin(8)$, $Spin(10)$ , $Spin(12)$ or $E_6$, and we argue in the same way
 as above. 

 To prove {\it (c)}, note first that the only ``interesting'' possibilities
for the simple factors of (the universal covering of) the centralizer of an 
element $g \in E_8$ 
(the  factors which are not of $A$ or $C$ type ) are the following:
  {\it (i)}  $E_7$, $E_6$, $Spin(12)$,  $Spin(10)$, $Spin(8)$; 
{\it (ii)} $Spin(16)$, and  there can be only one such ``interesting'' factor. 
Arguing as above , we show that any non--trivial triple of $G_{a_1}$ remains
non--trivial in $E_8$, hence $E_8$ has no minimal quadruples. In case 
{\it (i)}, $G_{a_1}$ contains no minimal $n$--tuples for $n \geq 4$ and
hence $E_8$ has no non--trivial $n$--tuples for $n \geq 5$. In the remaining
case {\it (ii)}, $G_{a_1} = Spin(16)/{\bf Z}_2$ contains a unique up to conjugacy
minimal quadruple which gives us the unique up to conjugacy
minimal quintuple in $E_8$ constructed in Example D3. 

Also $E_8$ contains no minimal $n$--tuples with $n > 5$ since $G_{a_1}$
contains no minimal $n$--tuples with $n \geq 5$ (see Example D2 and also the
final part of the proof). 

Consider finally the groups $Spin(N)$, $N \geq 15$. 
 Let us explain first why the reasoning used for the groups with $N < 15$ 
does not work in this case.  
The assumption that  $Spin(N)$ contains a non--trivial quadruple $(a_1, a_2,
a_3, a_4)$ allows one to conclude that the triple $(a_2, a_3, a_4)$ is 
non--trivial
in the centralizer $G_{a_1}$. If $G_{a_1}$ involves only one simple factor
admitting a non--trivial triple, we can argue as above to show that the
triple  $(a_2, a_3, a_4)$ remains non--trivial in $Spin(N)$, etc. But for
$N \geq 15$,
there is at least  one element $g \in Spin(N)$ such that 
the semi--simple part of  its centralizer
$G_g$ has {\it two} simple factors admitting a non--trivial triple. Let $a_1$
be such an element and let $K$ and $H$ be such factors. By Theorem 1, 
$K \equiv Spin(M_1)$ and $H \equiv Spin(M_2)$ with $M_{1,2} \geq 7$ and
$M_1 + M_2 \leq N$. 
A non--trivial
triple in $G_{a_1}$ can in this case be of different types. In can be 
non--trivial
in only one such factor and be trivial in another factor. In that case, the
triple $(a_2, a_3, a_4)$ can be presented  in the form (\ref{cantrip1}),
i.e. it remains non--trivial in $Spin(N)$.

 But a  triple which is non--trivial
 both in $K$ and in $H$ becomes trivial in the large group. Indeed, a 
non--trivial
triple in $K$ has the form $(a_K t_1^K,  b_K t_2^K, c_K t_3^K)$ where 
$(a_K, b_K, c_K) \in Spin(7) \subseteq K$ and 
$t_i^K$ lie on the maximal torus $T^K$ of the centralizer of the canonical
triple $(a_K, b_K, c_K)$ in $K$.
The same concerns  the triple in $H$.
One can choose now a maximal torus in $Spin(N)$ which involves $T^K,\ T^H$,
and also the torus formed by the generators like in Eq.(\ref{tormix}) which 
``mix'' two $Spin(7)$ factors embedding the canonical triples $(a_K, b_K, c_K)$
and $(a_H, b_H, c_H)$. 
 It is clear that this maximal torus contains  the triple $(a_2, a_3, a_4)$ 
which is thereby trivial and the quadruple $(a_1, a_2, a_3, a_4)$ is minimal.

In the range $15 \leq N < 31$, the $Spin(N)$ groups admit minimal quadruples,
but  not minimal quintuples, 6--tuples etc. The reason is that the centralizer
of any element of such $Spin(N)$ group contains at most one simple factor
admitting a minimal quadruple and the reasoning which proves the clause {\it
(b)} above works in this case. Starting from $Spin(31)$, the elements whose
centralizer involves two simple factors admitting a minimal quadruple
appear. Let $a_1$ be such element and let $(a_2,a_3,a_4,a_5)$ presents a 
minimal quadruple in each such factor. Arguing as above we show that the
quadruple $(a_2,a_3,a_4,a_5)$ is trivial in $Spin(N)$ and the quintuple
$(a_1,\ldots,a_5)$ is minimal. 
Likewise, the groups $Spin(N)$ admit minimal 6--tuple starting from
$N = 63$, etc.  

For $N > 2^n$, minimal $n$--tuples form the moduli space  (\ref{mod4Spin}).

{\bf Remark D1}. The minimal quintuple of $E_8$ is equivalent to that 
constructed in Ref.\cite{Alex}, which defines via the adjoint representation
a ${\bf Z}_2^5$ -- gradation of the Lie algebra of $E_8$ such that the eigenspaces
attached to all non--trivial characters  of  ${\bf Z}_2^5$ are 8--dimensional
(and that with a trivial character is trivial). Likewise, the isolated triples
consisting of $m$--exceptional elements with prime $m$
of $G_2$, $Spin(7)$, $Spin(8)$, $F_4$, $E_6$, and $E_8$ define, respectively,
${\bf Z}_2^3, {\bf Z}_2^3, {\bf Z}_2^3, {\bf Z}_3^3, {\bf Z}_3^3$, and
${\bf Z}_5^3$ -- gradations of the Lie algebras such that the eigenspaces 
attached to all non--trivial characters are 2--,   3--,  4--,  2--,  3--,
 and  2--dimensional, respectively \cite{Alex}.

Similarly, the isolated quadruples for $Spin(15)$  [resp. $Spin(16)$]
 define 
${\bf Z}_2^4$--gradations  of the corresponding Lie algebra 
with eigenspaces of dimensions 7 
(resp. 8), the isolated quintuples for $Spin(31)$ [resp. $Spin(32)$]
 define 
 ${\bf Z}_2^5$--gradations of the corresponding Lie algebra 
with eigenspaces of dimensions 15 
(resp. 16), etc. 

Let us illustrate  the $Spin(7)$ example. As was noted in
Ref.\cite{KRS}, the standard generators
$T_{ij} = i\gamma_i \gamma_j/2$ of the $Spin(7)$ Lie algebra  either
are invariant or change sign under conjugation by the elements of the triple
(\ref{Spin7}). The 
${\bf Z}_2^3$--gradation subdivides 21 generators of $Spin(7)$ in 7 classes 
involving each 3 commuting generators as in Table 9.

\begin{table}
\label{gradation}

\begin{tabular}{||c|c|c||} 


The generators & change sign
when conjugated by   & are invariant  when conjugated by  \\ \hline
$T_{15}, T_{26}, T_{37}$ & $\Omega_1$  & $\Omega_2,\ \Omega_3$
\\ \hline
$T_{13}, T_{24}, T_{57}$ & $\Omega_2$  &  $\Omega_1,\ \Omega_3$\\ \hline
$T_{12}, T_{34}, T_{56}$ &  $ \Omega_3$  & $\Omega_1,\ \Omega_2$ \\ \hline
$T_{17}, T_{35}, T_{46}$ & $\Omega_1,\ \Omega_2$  &   $\Omega_3$\\ \hline
$T_{16}, T_{25}, T_{47}$ &  $\Omega_1,\ \Omega_3$  &  $\Omega_2$\\ \hline
$T_{14}, T_{23}, T_{67}$ &  $\Omega_2,\ \Omega_3$  &  $\Omega_1$\\ \hline
$T_{27}, T_{36}, T_{45}$ &  $\Omega_1, \ \Omega_2,\ \Omega_3$ &  ---  \\ \hline

\end{tabular}

\caption{${\bf Z}_2^3$--gradation of $Spin(7)$.}

\end{table}

{\bf Remark D2}. In a connected, but not simply connected $G$, the 
classification of  commuting pairs follows from Theorem A1, and the
classification of minimal commuting $n$--tuples for $n > 2$ follows from
that in the simply connected cover of $G$.
If $G$ is not connected, non--trivial "1--tuples" appear which are 
classified by Gantmacher's theorem \cite{G}. Thereby, we have the 
classification of minimal commuting $n$--tuples in any compact Lie group
$G$.

\vspace{.3 cm}

The problem of classification the non--trivial triples is related to the 
problem of counting the vacuum states in 4--dimensional supersymmetric 
Yang--Mills theory. We are not aware about possible physical applications
of the problem of classification of minimal $n$--tuples, but
it does not mean that such applications do not exist.

\end{document}